\documentclass[acmsmall]{acmart}

\AtBeginDocument{%
  \providecommand\BibTeX{{%
    \normalfont B\kern-0.5em{\scshape i\kern-0.25em b}\kern-0.8em\TeX}}}

\acmJournal{PACMHCI}
\acmYear{2023} \acmVolume{7} \acmNumber{CSCW1}
\acmArticle{133} \acmMonth{4} \acmPrice{15.00}
\acmDOI{10.1145/3579609}

\usepackage{soul}
\usepackage{pifont}
\usepackage{xspace}
\usepackage{subfigure}
\usepackage{tabularx}

\renewcommand{\paragraph}[1]{\vspace{5pt}\noindent\textbf{#1\quad}}

\newcommand{\RQone}{What {are} hate raids: how are they orchestrated and who do they target?}
\newcommand{\RQtwo}{How do hate raids affect members of targeted groups?}
\newcommand{\RQthree}{How did different groups of stakeholders respond?}

\begin{document}

\setcopyright{acmlicensed}
\acmJournal{PACMHCI}
\acmYear{2023} \acmVolume{7} \acmNumber{CSCW1} \acmArticle{133} \acmMonth{4} \acmPrice{15.00}\acmDOI{10.1145/3579609}
\title{Hate Raids on Twitch: Echoes of the Past, New Modalities, and Implications for Platform Governance}

\author{Catherine Han}
\email{cathan@stanford.edu}
\affiliation{%
  \institution{Computer Science Department, Stanford University}
  \city{Stanford}
  \country{USA}
}

\author{Joseph Seering}
\email{jseering@stanford.edu}
\affiliation{%
  \institution{Computer Science Department, Stanford University}
  \city{Stanford}
  \country{USA}
}

\author{Deepak Kumar}
\email{kumarde@stanford.edu}
\affiliation{%
  \institution{Computer Science Department, Stanford University}
  \city{Stanford}
  \country{USA}
}

\author{Jeffrey T. Hancock}
\email{hancockj@stanford.edu}
\affiliation{%
  \institution{Communication Department, Stanford University}
  \city{Stanford}
  \country{USA}
}

\author{Zakir Durumeric}
\email{zakird@stanford.edu}
\affiliation{%
  \institution{Computer Science Department, Stanford University}
  \city{Stanford}
  \country{USA}
}

\renewcommand{\shortauthors}{Catherine Han et al.}

\begin{abstract}
In the summer of 2021, users on the livestreaming platform Twitch were targeted by a wave of ``hate raids,'' a form of attack that overwhelms a streamer's chatroom with hateful messages, often through the use of bots and automation. Using a mixed-methods approach, we combine a quantitative measurement of attacks across the platform with interviews of streamers and third-party bot developers. We present evidence that confirms that some hate raids were highly-targeted, hate-driven attacks, but we also observe another mode of hate raid similar to networked harassment and specific forms of subcultural trolling. We show that the streamers who self-identify as LGBTQ+ and/or Black were disproportionately targeted and that hate raid messages were most commonly rooted in anti-Black racism and antisemitism. We also document how these attacks elicited rapid community responses in both bolstering reactive moderation and developing proactive mitigations for future attacks. We conclude by discussing how platforms can better prepare for attacks and protect at-risk communities while considering the division of labor between community moderators, tool-builders, and platforms.
\end{abstract}



\maketitle

\section{Introduction}
\vspace{1mm} 

\noindent\fbox{%
    \parbox{0.96\columnwidth}{%
        \textbf{Content Warning}: This paper studies hateful online content. When
        necessary for clarity, this paper directly quotes user-generated content
        that contains offensive/hateful speech, profanity, and other potentially
        triggering content.
    }%
}
\vspace{2mm}

Livestreaming platforms have boomed in popularity in recent years and become a major part of many users' Internet experience. The livestreaming industry saw a 45\% uptick in viewership between March and April 2020~\cite{stephen202NEWSlivestreaming}, likely in part due to the COVID-19 pandemic. Twitch is a popular livestreaming platform, and much like any other rapidly growing online platform, its communities have suffered from hate and harassment. July to October 2021 marked an intense period of harassment on Twitch with many streamers experiencing a surge of ``hate raids.'' In the most common form of hate raid, a streamer's chatroom is overwhelmed by a rapid influx of Twitch accounts posting hateful messages. Because of their dissatisfaction with Twitch's handling of hate raids and poor treatment of marginalized-identity streamers, these streamers and their communities came together, gathering resources, developing tools and strategies to protect themselves, and organizing a major protest~\cite{grayson2021NEWStwitchdobetter,parrish2021NEWSprotests}. This series of events reflected frustration within the community---particularly from minority streamers---toward Twitch and its perceived inaction on issues of trust, security, and safety on the platform.

In this paper, we investigate the nature of hate raids on Twitch, how they affected vulnerable communities, and how stakeholders reacted to hate raids. We combine an at-scale measurement of the hate raid phenomenon across 9,664~popular channels' chats on Twitch with interviews of seven LGBTQ+ and/or Black Twitch streamers. In addition, we interview two Twitch users that developed third-party moderation tools in response to hate raids. In our analysis, we explore the following three research questions:

\paragraph{RQ1: \RQone}
We first seek to detail the fundamental characteristics of hate raids.\footnote{Some news reports stated that these raids began as an abuse of a built-in ``raiding'' feature originally intended to help grow a sense of community~\cite{pandey2021NEWShateraids}, but we did not find direct evidence of this in our dataset.} Our measurement of hate raids across 9,664~popular channels on Twitch reveals that 98\% of hate raid messages consisted of identity-based attacks. However, while the \textit{content} of these attacks was mostly anti-Black or antisemitic, the raids themselves selected targets indiscriminately with respect to streamer identity. These hate raids blurred the line between what prior work called ``trolling'' or disruptive behavior~\cite{phillips2015we} and networked harassment~\cite{marwick2021networked}. To better understand how attackers selected their targets, we examined Twitch's streamer tags---a feature streamers use to categorize themselves and their community. Among streams that use tags, we find evidence that attackers may have leveraged these tags to discover and attack marginalized-identity streamers: particularly with Black, African American, and LGBTQ+ tags. 

\paragraph{RQ2: \RQtwo}
Because a quantitative perspective on hate raids cannot fully depict the lived experiences of targeted community members, we interviewed seven Black and/or LGBTQ+ streamers on Twitch about the impact of these attacks. Through these interviews, we find that the perspectives of targeted streamers aligned with mainstream media portrayals of these attacks: hate raids are seen as highly-targeted attacks often persecuting Black and LGBTQ+ communities on Twitch. While identity-based attacks have always plagued these at-risk communities online, streamers found that this wave of hate raids was distinct in its highly-targeted nature and the persistence of its perpetrators. Furthermore, we find that the community saw hate raids as one piece of a larger campaign of harassment, often involving other platforms and in some cases extending into more extreme offline experiences (e.g., involving law enforcement, swatting\footnote{A harassment tactic that involves calling emergency services or police to a target's residence}).

\paragraph{RQ3: \RQthree}
To better understand the different ways community members and Twitch responded to hate raids, we further draw upon data from interviews with streamers and bot developers. We observed that streamers largely turned to their community and third-party bot developers for moderation, emotional, and technical support against hate raids. Volunteer bot developers created tools adopted by tens of thousands of streamers who felt that they might be targeted. These developers worked to constantly update their tools throughout the hate raid period, as the sophistication of hate raids evolved in response to developers' efforts to combat these attacks. In addition to an influx of support via resource aggregation, tool development, and volunteer moderation, the community rallied together for a social movement and virtual walkout to raise awareness for their longstanding frustrations with Twitch. While attitudes toward the degree of success of these movements varied among our interviewees, these community-driven movements gained attention and impacted overall platform engagement.

Our mixed-methods approach to understanding hate raids provides the following three primary research contributions: 
\begin{enumerate}
    \item  We characterize a novel form of long-term harassment campaigns on Twitch; not only do we observe that hate raids leverage the real-time nature of livestreaming platforms, but we also find that they exploit automation to select targets and amplify their attacks.
    \item We observe that the content and orchestration of hate raid messages indicate a dual motivation: first, hate-driven and second, attention-seeking, consistent with prior research into networked harassment and subcultural trolling.
    \item We find that members of these targeted communities, unhindered by the frictions platforms face when developing new features and policies, rapidly assembled high-quality resources and produced technical tools to address their needs and the limitations of Twitch's response.
\end{enumerate}

\noindent
Grounded in our data, we conclude by discussing the implications of our findings for livestreaming platform design and the broader community. We argue that platforms and researchers must proactively consider the unique experiences of targeted communities online, the dependency on and potential for community-based moderation and tool-building, and the range of motivations behind the actors coordinating hate-based attacks.


\section{Related Work}
This paper builds on three key bodies of related work. First, we review literature on morally-motivated networked harassment \cite{marwick2021networked} and subcultural trolling \cite{phillips2015we}, and we identify characteristics of each that hate raids share. Second, we review online hate-based attacks documented in the literature, situating hate raids within taxonomies of their characteristics. Finally, we discuss ties to literature on volunteer moderation and coordinated action, identifying connections between hate raids and crisis informatics literature and highlighting how users' responses to hate raids parallel responses to natural disasters and other crises.

\subsection{Harassment and ``trolling'' in online spaces}
In this paper, we situate the Twitch hate raids within prior work that discusses online harassment and ``trolling.'' Definitions for both of these terms have varied widely; for example, trolling has been defined as broadly as ``behavior that falls outside acceptable bounds defined by [...] communities'' \cite[p.~1]{cheng2017anyone} and as specifically as in Phillips' description of ``subcultural trolling'' as a nuanced cultural phenomenon with historical and moral roots \cite{Phillips2011memorial, phillips2015we}. Similarly, Marwick identified more than ten different types of behaviors listed under the umbrella term of ``online harassment'' in prior work \cite[p.~2]{marwick2021networked}. We operate under the definitions of the two terms provided by Marwick and Phillips, and we focus on the form of harassment that Marwick terms ``networked harassment,'' where an individual is harassed by many people connected by social media. 

Note that, subsequent to her original publications on subcultural trolling, Phillips wrote about the dangers of referring to something as ``just'' trolling \cite[p.~2]{phillips2019exclusionary}. While in this paper, we compare aspects of hate raids to aspects of Phillips' characterization of subcultural trolling, this should not be construed to mean that hate raids are ``just'' trolling by any means; they cause real harm to targets that should not be taken lightly. Moreover, these attacks occurred in the context of a long history of racist, sexist, and transphobic behaviors in online spaces that have been especially prevalent in online gaming spaces \cite{fox2017womens, gray2017urban}. These behaviors have forced targeted users to hide their identities or even to withdraw from online spaces entirely \cite{vitak2017identifying, fox2017womens, cote2017strategies, scheuerman2018safespaces}.

\subsection{Characterizing hate-based attacks}

Thomas et al. \cite{thomas2021hatesok} identify three axes on which hate-based attacks can be classified: 

\begin{enumerate}
    \item The \textit{Audience} exposed to the attack, which can include the target and/or a different audience.
    \item The \textit{Medium} through which the attacker reaches a target, which frequently includes media such as text, images, or video. 
    \item The \textit{Capabilities} that are required for the attack to succeed: whether the attack requires deception of an audience and/or a third-party authority, whether it requires amplification, and whether it requires privileged access to information, an account, or a device.
\end{enumerate}

In the context of online hate and harassment behaviors, the most similar to hate raids  is ``brigading,'' where a single target (e.g., a YouTube video or Twitter account) is simultaneously attacked by a semi-coordinated set of antagonistic users. For example, 4chan users often coordinate to target YouTube videos that they are ideologically or otherwise opposed to~\cite{mariconti2019you}; Reddit users have previously, in large groups, entered other community spaces to harass and intimidate other subreddits~\cite{datta2019extracting}; Zoom users have leveraged legitimate insider access to join online meetings to disrupt and harass the other participants, otherwise known as ``Zoombombing''~\cite{ling2021zoom}.

Of the above criteria, the \textit{medium} through which hate raids took place is primarily text, though in some cases other media on external platforms were involved. As we discuss later in this work, they required an \textit{audience} that included both the target and a wider array of viewers. In some cases, the attacks included revealing personal information of targets (``doxxing''), and they benefited greatly from amplification.

However, as we discuss in Section~\ref{sec:characterization}, these attacks had a number of other attributes worth mentioning. For example, the \textit{capabilities} required for this attack included that they were heavily automated and occurred over a significant period of time (several months), hearkening to more traditional cybersecurity attacks, such as Distributed Denial-of-Service (DDoS)~\cite{moore2006inferring} and for-profit spam and scam campaigns~\cite{kanich2008spamalytics}. Though Zoombombing often operates under a notion of the infiltration of a private meeting, public Twitch streams share the capability of seeing the reactions and impact of the attack in Zoombombing attacks. Therefore, we draw upon prior work in the cybersecurity space to structure our understanding of abuse executed en masse via illegitimate accounts. Contextualizing the hate raids on Twitch through both a lens of subcultural trolling and morally-motivated networked harassment and a traditional cybersecurity lens better frames the underlying motivation and tactics of these activities.

\subsection{Volunteer moderation, coordinated action, and crisis informatics}

Prior work examining platform governance and volunteer labor in online social spaces has highlighted a variety of dynamics that inform our analysis of hate raids. While Twitch is a multi-modal platform incorporating text-based chat, video, and audio, the phenomenon of hate raids echoes the moderation challenges discussed by Jiang et al. for voice-based communities~\cite{Jiang2019voice}, as both Twitch and Discord share ephemeral and real-time components of user interactions. Additionally, we discuss the experience of hate raids and the resulting mobilization of less visible streamers on Twitch and members of marginalized communities on the platform more broadly. Prior work details the obstacles that such communities in particular face with regards to platform visibility and accountability~\cite{thach2022invisible}, further contextualizing the friction we observe between Twitch and its users. Several examples~\cite{Seering2017shaping,seering2022pride,cai2022coordination} in the literature emphasize the importance of volunteer labor in these communities, reporting that volunteer moderators on livestreaming platforms --- both individually and in collaboration --- have the capacity to effectively and quickly address norm-violating behaviors. In Section~\ref{sec:response}, we discuss the impact of community moderation and community-developed automated moderation tools, adding to conversation in prior work that has raised questions surrounding platform governance and the distribution of labor in content moderation~\cite{roberts2016commercial,seering2019moderator,kiene2016surviving,chandrasekharan2019crossmod}. 

As we detail below, one of the core characteristics of users' responses was collective action to create tools and aggregate informational resources. A small number of examples of collective action to counter harassment at this scale have been documented in social computing literature. Blackwell et al. reported on ``HeartMob,'' a platform where users can submit reports of being harassed and volunteers will provide support --- supportive messages, help with reporting harassment, and/or help documenting abuse \cite{Blackwell2017classification}. On a much smaller scale, Mahar, Zhang, and Karger's ``Squadbox'' allowed users to coordinate trusted friends to help shield them from harassment via email \cite{mahar2018squadbox}. A small body of work from the early-mid 1990s \cite{Dibbell1993rape, mackinnon1997virtual, Smith1999communities} and early 2000s \cite{herring2002} also documented individual cases of harassment and communities' discussions about how to respond. 

A broader related body of work, situated in part in CSCW literature, comes from the field of crisis informatics \cite{palen2007crisis, palen2016newdata}. Though this field has largely focused on responses to offline crises (e.g., natural disasters \cite{soden2014crowdsourced, soden2016infrastructure, metaxa2018hurricane, zade2018crisisresponse}, terrorist attacks and mass shootings \cite{castillo2016big, cheong2011terrorism, palen2007crisis}, and in some cases ongoing violent conflict \cite{monroyhernandez2013correspondents, semaan2011breakdowns}), many of the core principles are also mirrored in responses to hateful attacks based on social media. As we discuss in Section~\ref{sec:response}, we observe many of the same behaviors in our research on Twitch hate raids that occur during natural disaster response. Per this literature, we have organized our results to address questions about crisis response that parallel questions commonly asked in crisis informatics literature.

\section{Methods}
We examine broad patterns in hate raids and common themes in individual messages, and we complement this analysis with insights from interviews with impacted individuals. In this section, we describe the methodologies of our (1) large-scale collection and analysis of Twitch chat messages, moderation actions, and channel attributes collected from 9,664~channels from September 2 to September 16, 2021, (2) interviews with seven Black and/or LGBTQ+ streamers, and (3) interviews with developers of two third-party Twitch moderation bots that were widely deployed in response to hate raids. 




\subsection{Twitch Chat Data Collection}

To understand how hate raids impacted high-visibility streams on Twitch, we generated a corpus of channels to gather messages from. We used Twitch's API to pull information about online streamers ordered by their current number of viewers, from high to low. We pulled this data every hour for a week from May 4 to May 11, 2021 to compute an average number of viewers per stream when the channel was live. For our corpus, we only considered channels that had an average of at least 100~viewers each time they streamed and that also streamed at least three times over the course of a week. 

We continuously gathered data from the channels on this list for two weeks in September, from September 2 to September 16, 2021, during which time many hate raid attacks occurred. Each channel on Twitch has an associated chatroom built on Internet Relay Chat (IRC) protocols. When connecting to each channel's chat, we sent requests for information about the channel's chatroom modes---unique chat, subscribers-only mode, and slow mode.\footnote{\url{https://help.twitch.tv/s/article/chat-commands}} We also sent requests for \textit{command} and \textit{membership} capabilities, which allow us to identify the usage of certain moderation and room state commands and to determine when users joined or left chat; the \texttt{CLEARCHAT} command indicates that all of a specific users' messages were purged from the chat, often as a result of a moderation action, like a timeout or a ban, 
while the membership capability reveals when specific users are joining and leaving the chat. In total, we collected 244,738,672~messages. For each message that was sent, we collected various pieces of metadata to contextualize it: what channel it was sent in, the account that sent the message, the text content of the message, the timestamp of when it was sent to the chat, the status of the chatroom (e.g., if it was in ``slow mode''), and basic, publicly-visible information about the account that sent the message (e.g., if the account is a subscriber, follower, or moderator of the channel it is participating in). All data that was collected for this portion of this study was public to any user viewing the stream.

\subsection{Detecting Hate Raids}
\begin{figure}
\includegraphics[width=0.99\linewidth]{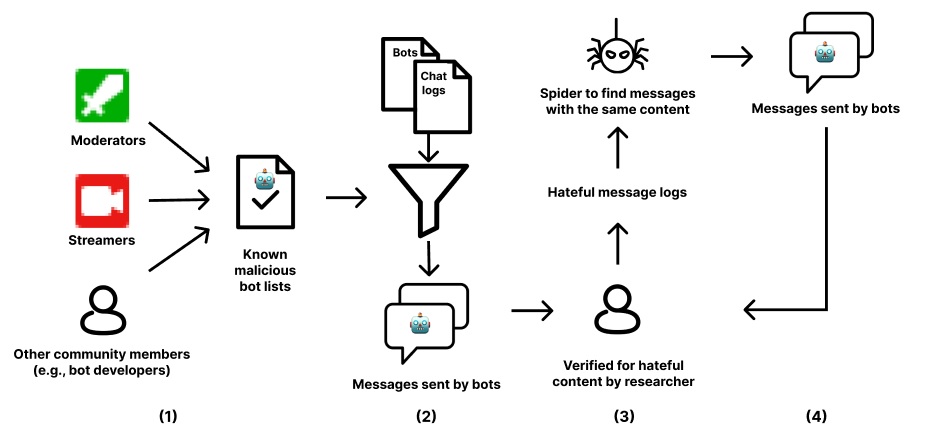}
\caption{Compiling hate raid logs. We collected the hate raid logs in a series of four steps: (1) we began with a list of known malicious bot account names collected by moderators, streamers, and other community members on Twitch; (2) we used this to filter the chat logs for messages sent by malicious bots; (3) after confirming the contents of these messages are hateful, we used this seed set of messages to spider for similar content being sent by accounts not already in this list; (4) we iterated with this notion of message content similarity until no new bot accounts were detected.}
\label{fig:bot_crawl}
\end{figure}

We started with a collection of 1,319,890~likely malicious bot accounts curated by and shared among the Twitch community so that streamers could proactively ban and block these accounts from participating in their chats. We searched our Twitch chat dataset for messages sent by these accounts, creating a seed set of messages from 516 of these likely bot accounts. We then used approximate string matching computed using the Levenshtein distance with a threshold of 95\% similarity to find messages with the same content despite some evasion techniques used by hate raid attackers, such as prepending randomness to the same message contents across different accounts. We continue this process of finding approximate message content until no new messages were discovered. Through this method, we found matching message contents found by an additional 1,067 discovered bot accounts for a total of 1,583 bots participating in hate raids (Figure~\ref{fig:bot_crawl}). We then determined hate raid events to be windows of time where bot accounts in our dataset were seen sending messages within two minutes of prior messages sent by bots. We restricted this window to a short interval because raiding behavior (both benign and malicious) often involves an influx of similar messages sent across different accounts within a short period of time.

\subsection{Streamer and Third-Party Bot Developer Interviews}
We conducted semi-structured interviews with seven Twitch streamers who identified as Black and/or LGBTQ+ and with two Twitch users who created third-party moderation bots to combat hate raids. These interviews were conducted from early October through mid-November 2021, shortly after the major spike in hate raids in late September. Interviews lasted between 20 minutes and one hour, with length varying based on participants' exposure to hate raids, their roles within the community, and their knowledge of moderation tools. We recruited participants from lists of streamers who had previously participated in visible roles during LGBTQ+ focused events on Twitch, including featured streamers during Pride Month, streamers who were reported in news articles as having been heavily targeted by hate raids, and streamers who actively participated in hate raid-focused conversations in both public and semi-private spaces dedicated to hate raid responses. We recruited specifically from Black and LGBTQ+ streamers because these were the groups at the center of discourse surrounding hate raids and were the most visibly targeted. Interview questions focused on the same topics as the research questions, with a full list of primary questions presented in Appendix ~\ref{appendix:interview_questions}. Due to the open-ended, semi-structured nature of these interviews, we asked additional follow-up questions when relevant.

We do not report identity characteristics for each streamer individually because doing so might identify them, but the following are aggregated, self-reported demographic categories: four streamers identified as Black, two as Hispanic, and one as white. Five identified as women, one identified as nonbinary, two identified as transgender, two identified as queer, and one also identified as aromantic and asexual. Some interviewees identified with more than one of these categories. The bot developer interviewees both identified as white, male, and heterosexual. 

Following interview completion and transcription, interview text was separated into chunks. Each chunk contained a single idea, which ranged in length from several words to several sentences using a variant of the method described in Creswell \cite[p.~ 86--89, 184--185]{creswell2013qualitative}. These chunks were each given category labels, which included categories such as ``Frequency of hate raids experienced,'' ``Streamers' short-term responses to hate raids,'' and ``Social support received by streamers.'' The full codebook is included in Appendix~\ref{interview_codebooks}. Initial category labels were defined by the research questions, but labels were iteratively added to the codebook when a chunk did not fit any existing labels. A Cohen's Kappa statistic was calculated to determine inter-rater reliability, with the final round of coding achieving a Cohen's Kappa of 0.91. The results of this analysis are summarized by category label in Section ~\ref{sec:response}.

\subsection{Ethical Considerations}
We gathered data from 9,664 different Twitch channels, each with at least 100 viewers on average. Even though chat data from all channels on Twitch is publicly viewable, we elected to restrict the scope of our analysis to this set of larger channels to protect any assumption of privacy that smaller channels and their communities might have; channels with regular audiences of 100 or more viewers represent an exceedingly small proportion of Twitch channels overall---in May 2021, nearly 99\% of streams had fewer than 50 average concurrent viewers~\cite{cooney2021NEWSviewers}. This restriction applies a significant limitation to our quantitative analysis, as we cannot draw conclusions regarding hate raid messages sent to smaller channels, but we believe that the ethical considerations in respecting privacy justify this limitation. The final list contains 9,664 active channels that match these criteria. In the chat data we collected, we took precautions to minimize the risk of inadvertently affecting communities: our script did not send any messages or interact with the chat, and we did not attempt to de-anonymize the involved accounts.

Furthermore, to gather our qualitative data, we interviewed members of Black and/or LGBTQ+ communities concerning their experiences with hate raids. Because of the sensitive nature of this research, participants were notified of the full purpose of the interview in advance, as well as what types of questions would be asked. Additionally, we reminded participants both on the consent form and at the beginning of the interview that they could decline to answer any questions or stop at any time. Interviews typically lasted between 20 to 60 minutes, and participants were compensated with an Amazon gift code for \$15 or local currency equivalent. To protect participants' anonymity, we have removed any potentially personally identifiable information from their quotes. This work was approved by the Stanford University Institutional Review Board (IRB).

\section{Results}
We measure hate raids across the platform and present our findings of their quantitative characteristics below (Section \ref{sec:characterization}). We pair these measurements with a synthesis of the qualitative perspectives of streamers from at-risk communities and community bot developers on the responses of different stakeholders (\ref{sec:response}).

\subsection{Characterizations of Hate Raids}
\label{sec:characterization}

Mainstream news outlets characterized the hate raids during late summer of 2021 as targeted, bot-mediated abuse often aimed toward marginalized streamers~\cite{danastasio2021NEWShateraids,grayson2021NEWShateraids,pandey2021NEWShateraids}.
We find two forms hate raids: first, a broad, scattershot form of hate raids akin to classic subcultural trolling~\cite{phillips2015we} that incorporates racist and antisemitic elements, and second, hate raids that targeted specific streamers based on their identities.

\subsubsection{Quantitative Perspectives}\label{sss:quantitative}
We first sought to understand what hate raids looked like quantitatively across the platform.\footnote{Note that, as discussed above, we focus here on within-chat hate raids rather than on forms of follow-botting that were sometimes included under the umbrella term of hate raids.}
To achieve this, we characterized hate raids observed in a corpus of 244M~messages across 9,664~channels collected during a 14-day period from September 2 to September 16, 2021.
Of these messages, 2,947~messages were identified as being part of hate raids through the methods discussed above. We observed 60~hate raid attacks in 57~unique channels---three of these channels were hate raided twice on separate occasions.

\begin{figure}
\includegraphics[width=0.6\linewidth]{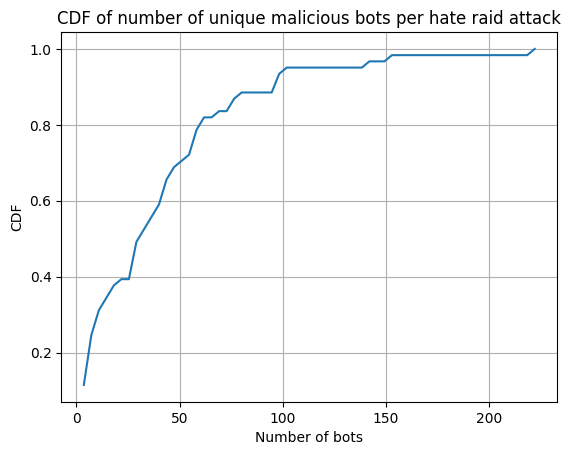}
\caption{CDF of the number of unique bots that participated in an instance of a hate raid. We find that the median bot count was 32, demonstrating the typical scale of these attacks.}
  \label{fig:num_bot_accounts}
\end{figure}

\begin{figure}
\includegraphics[width=0.6\linewidth]{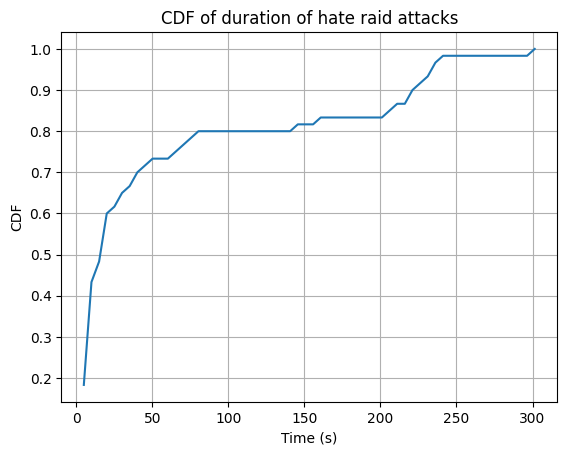}
\caption{CDF of the duration of hate raid instances in seconds. The hate raids we observed were largely a short burst of hateful messages sent in close succession, often within the span of seconds or minutes.}
\label{fig:raid_duration}
\end{figure}

\begin{figure}
\includegraphics[width=0.6\linewidth]{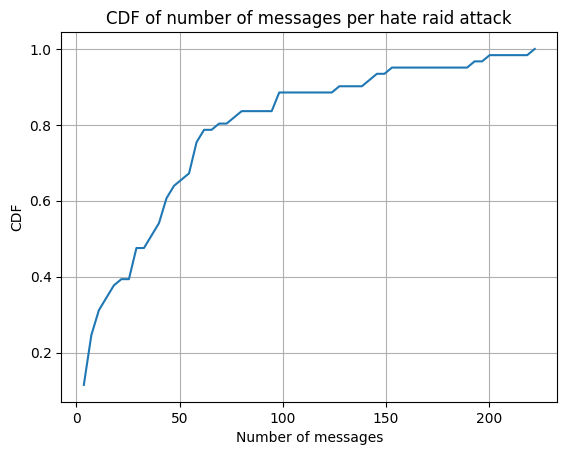}
\caption{CDF of the number of messages sent per hate raid instance. The median message count is 36, showing the usual scale of the observed hate raids. Note that this number reflects how many messages actually appeared in the stream chat; reactive moderation actions taken by the streamer and/or their volunteer moderators may have prevented bots from sending additional messages.}
\label{fig:raid_message_count}
\end{figure}

\paragraph{Technical Characteristics}
We find that 50\% of channels that were hate raided had at least 32~bot accounts involved in the attack (Figure~\ref{fig:num_bot_accounts}). Some channels, however, experienced attacks with an acutely large number of bots. For example, one channel received hate raid messages from 222~unique bot accounts. We found that on average, there were 48~messages per raid. These messages were typically sent in close succession to one another. In the majority of raids, all of the messages were sent in less than 16 seconds (Figure~\ref{fig:raid_duration}), though a smaller proportion of raids lasted for minutes. Most messages were sent from unique bot accounts, with 302 bots (19.1\%) sending more than one message in the same raid; even when these bots did send more than one message, the median number of messages sent by a single bot was two (Figure~\ref{fig:raid_message_count}). 

\begin{figure}
    \centering
    \includegraphics[width=0.6\linewidth]{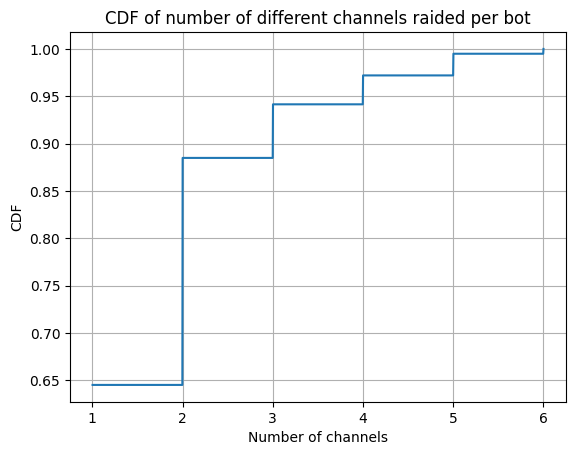}
    \caption{CDF of the number of different channels a bot account hate raids. Almost 65\% of the bot accounts we identified were found in only one channel, implying that the majority of these bots were created for a single use in our observation period, though it is possible that these bots were used additional times in channels that were not in our sample.}
    \label{fig:num_channels_raided}
\end{figure}



Overall, bots appear to have been largely throwaway, single-use accounts often created for the purpose of enacting these hate raids. The usernames of the bots we observe in our dataset were predominantly (99.6\%) strings of letters and numbers that appear to have been automatically generated (e.g., \texttt{y9y7n18r0g6raem}). Of the bots we detected sending messages ($N=1,583$), many (25\%) of the bots were created within a two-day window of their first use in our dataset. While these bots appear to have been created for use in a single hate raid, we find that 3\% were made far in advance of their first use---these are accounts created at least several weeks before observed chatting in our corpus. This trend toward many single-use accounts likely controlled and created by a single entity echoes the literature in ``sockpuppetry,'' where accounts are created and controlled by a ``puppetmaster'' for engaging in deceptive behavior influencing the surrounding community~\cite{kumar2017sockpuppet}. Likewise, the small proportion of aged accounts may indicate what prior work defines as ``zombie'' accounts, which are ones that are created ahead of time but are dormant for a long period or indicate benign account compromise~\cite{elmas2022astroturf}.
    
We found limited evidence of bot account reuse; the majority of bots were only observed participating in a single channel within our sample, and even then, most bots sent only one message (Figure~\ref{fig:num_channels_raided}). We also found that slightly more than 20\% of the known bots sent hate raid messages in at least two different channels. As of the time of this analysis, there were two ways to sign up for a Twitch account: (1) through e-mail and (2) through phone number. The cost of creating Twitch accounts could thus be as low as the cost of creating e-mail accounts. During this period, Twitch did move quickly to disable accounts that participated in malicious activity. When querying Twitch's API to understand the ages of these accounts, we successfully fetched information for only 33 (4\%) of the bot accounts in our dataset because the malicious accounts had been disabled by the time we queried the API for their information. 

Only three of the 57 targeted channels experienced a hate raid attack more than once, meaning that our corpus contains 60 observed attacks.
Two of these channels experienced two attacks in close succession---within 30 minutes of the first attack. Additionally, both channels experienced a similar pattern where, between the two attacks, the shared text of the messages spammed by different bot accounts in the same raid changed. For instance, in one of the channels, the message content spammed in the first attack was a violent, anti-Black racist statement mocking Twitch community efforts to organize against hate raids. Just 27 minutes later, a second attack began with accounts spamming a different anti-Black racist message. We note that, while both raids were anti-Black in nature, the target of this raid was white.

\paragraph{Degree of Targeting}
Because of the tendency toward identity-based attacks we observed in the hate raid messages, we next more closely examined the contents of these messages for a semantic understanding of hate raid targeting. Although the bursty messaging pattern with identical message content that we observed in hate raids may appear similar to other, more benign behaviors on Twitch (e.g., chanting\footnote{An experimental feature introduced by Twitch in May 2021 that allows streamers and their moderators to ``suggest'' messages to be duplicated or ``chanted'' by other users.} and pro-social raiding), the content of hate raid messages distinguishes them as clearly malicious.
We find that the hate raids in our dataset spanned several different kinds of hate---most often identity-based---and weaponized these hateful ideologies via graphic and threatening language.

\begin{table}
\small
\setlength{\tabcolsep}{1.5pt}
\centering
\begin{tabularx}{.95\columnwidth}{XX}
\toprule
Hateful Ideology & Meaning\\
\midrule
Antisemitic            & Demonstrating prejudice toward Jewish people \\
Anti-Black              & Demonstrating prejudice toward Black people \\
Anti-Trans              & Demonstrating prejudice toward transgender people \\
Individual streamer    & Harassing a particular streamer or individual (not necessarily the streamer whose channel the message is sent in)\\
\bottomrule
\end{tabularx}
\begin{tabularx}{.95\columnwidth}{XX}
\toprule
Mode of Operation & Meaning\\
\midrule
Violent threat           & Threats of violence (describing explicit actions) \\
Known propaganda         & Using known hate symbols or references (e.g., most commonly excerpts from the Great Replacement or 1488*) \\
Direct attack            & Attacks that appear to directly address the streamer (e.g., attacks in the second person) or align with the streamer's identity \\
Fearmongering            & Inspiring fear or resignation by emphasizing the futility of counter-hate raid efforts \\
Weaponized emote         & Coopting Twitch emotes for harassment purposes (e.g., TriHard**) \\
Dehumanization           & Implications that a group of people is not human (e.g., comparisons with animals) \\
\bottomrule
\end{tabularx}
\vspace{3pt}
\caption{\textbf{Codebook for hate raid message content}---%
    Codebooks for hate raid message content separated into two axes: (1) hateful ideology and (2) the mode in which they were operationalized.}
{* 1488 is a pairing of two popular hate symbols, both regarding white supremacy and neo-Nazism.}

{** TriHard is a global Twitch emote depicting the face of a Black streamer, TriHex, and it has been used in the past to alienate Black streamers.}
\label{tbl:codebook}
\end{table}




\begin{table}
\small
\setlength{\tabcolsep}{1.5pt}
\centering
\begin{tabularx}{.95\columnwidth}{l|c|c|c|c|c}
\toprule
& Violent threat & Known propaganda & Direct attack & Fearmongering & Weaponized emote \\
\hline
Antisemitic         & 10 & 12 & 2 & 0 & 3 \\
Anti-Black           & 43 & 11 & 4 & 7 & 3 \\
Individual          & 6  & 0  & 0 & 3 & 0 \\
\bottomrule
\end{tabularx}
\vspace{3pt}
\caption{\textbf{Number of raids containing categories of operationalized hate}---%
    Anti-Black racism was the most common form of identity-based hate we identified, and it was most often operationalized through violent threats. We do not list anti-Trans attacks in this table because they were not present in our dataset of messages, though they may have been present in hate raids we were not able to capture.}
\label{tbl:code_grid}
\end{table}


To better characterize how hate was expressed in the raids, we categorized the content of the hate raid messages in our dataset. We evaluated each message along two axes: (1) what identities were attacked in the message, and (2) in what method this identity-based hate was operationalized.
Following best practices for grounded theory coding, two researchers agreed upon a master codebook (Table~\ref{tbl:codebook}) and independently coded 2,947 messages sent across 60 different attacks. The Kupper-Hafner agreement was computed to determine inter-rater agreement because some messages were assigned multiple labels, and the coding achieved an agreement of 0.85.
Ultimately, the researchers met to agree on the final codes.
We found that the most common category of hate expressed was anti-Black racism, which was present in nearly all of the hate raids in our dataset (59, 98.3\%).
We observed that anti-Black racism was most often operationalized through violent threats (43 of 59, 72.9\%).
We also noted that the content in hate raid messages was frequently an amalgam of hateful ideologies---for instance, while anti-Black racism is an explicit category of identity-based hate, messaging with anti-Black attacks often co-occured with QAnon propaganda (23 of 59, 39.0\%).
These ideologies often overlap with their hateful roots (e.g., white supremacy underlies anti-Black racism, antisemitism, and aspects of QAnon), but the way that these themes were presented together was typically disjointed, separated into different parts of a single message. For instance, the following message both expresses anti-Black racism and supports QAnon:
\begin{quote}
    \emph{``7i9nnde4k WayneLambright Legion | kiII -> bIacks | behead -> bIacks| <readacted> is a cloutchasing clown he isnt even hateraiding he is just following''}
\end{quote}


There are four clear parts to this message, each with a separate meaning. This message's pattern is a common one throughout our dataset; in this case, pipes (``|'') were used to delimit separate parts, as attackers presented several pieces of unrelated hateful content in one message, but other symbols (e.g., brackets, braces, ``==>'', etc.) were also used to separate or relate concepts. In this example, the first component promotes Wayne Lambright, who ran for president in the United States in 2020 on a campaign supporting QAnon, anti-Black racism, and pseudoscience. The second and third parts both express violent anti-Black threats.
Finally, the last piece is an attack on an individual streamer. We note that in the messages attacking this streamer, they are neither present as a user in the chat nor are they the streamer of the channel itself; these attacks attempted to harass this streamer through fabricating negative associations between them and hate raid orchestration.

\begin{table}[]
\begin{tabular}{lrrr}
                                \toprule
                                   & \multicolumn{3}{c}{\emph{Stream Tags}}  \\
                                   & LGBTQIA+ & Black+AfAm & No Tag   \\
                                \midrule
Hate raided streams (N=57)         & 13.5\%   & 9.6\%        & 5.2\%  \\
Not hate raided streams (N=9,664)  & 2  .4\%    & 0.2\%        & 15.6\% \\
                                \midrule
p-value                            & 0.01     & 0.01         & 0.11    \\
                                \bottomrule
\end{tabular}
\caption{\textbf{Streamer tags}---%
    The results of a two-sample proportion test of the self-assigned identity tags (e.g., LGBTQIA+, Black, African American) between the channels that were and were not hate raided. We find that both LGBTQIA+ and Black + African American tags were disproportionately represented among hate raided streams.}
\label{tbl:streamer_tags}
\end{table}

\paragraph{Streamer Tags}
Because the contents of these hate raid messages were largely rooted in anti-Black racism and antisemitism, we next investigated the use of Twitch tags as potential vectors for targeting.
When streaming on Twitch, streamers can choose to categorize their streams with ``tags,'' which are ways to publicly describe the stream for viewers to better search for streams of interest.
These tags are maintained by Twitch, but the list of available tags are updated based upon community feedback.
In May 2021, Twitch introduced over 350 opt-in tags for streamers to better categorize their channels into a particular community. These tags were largely identity-based, including ``gender, sexual orientation, race, nationality, ability, mental health, and more''~\cite{twitch2021NEWStags}.
However, some streamers feared that the same tags that were meant to increase visibility within a community could be abused to ``single out minority streamers,''~\cite{pandey2021NEWShateraids} and other members of these communities discouraged use of these tags as a preventative measure~\cite{horetski2022NEWShateraids} to hide themselves from potential attackers.
We observe 54 of 57 channels (94.7\%) tag themselves with at least one such category; however, we focus our attention on the tags that give insight into streamer identities (e.g., LGBTQ+, Black, etc.) because the messages consisted of identity-based attacks.

To better understand tags' potential usage as a mechanism for targeting marginalized communities, we performed two-sample proportion tests to compare the presence of these identity tags between channels that were and were not hate raided.
We find quantitative evidence that suggests that tags may indeed have been used to find targets for harassment at scale: both the LGBTQ+ and Black/African American tags were disproportionately represented ($p < 0.05$) in the hate-raided streams (Table ~\ref{tbl:streamer_tags}). We do \textit{not} find, however, any usage of the Jewish identity tag despite the heavy usage of anti-Semitic language in hate raid messages, and we note that the disproportionate representation of LGBTQ+ streamers deviates from the identities attacked in the \textit{contents} of the messages. 

In order to more fully understand the disparity between the identities of the targeted streamers and the identities attacked in the content of the message, we categorized the racial identities of (1) the streamers who were raided and (2) a random sample ($N=370$) of the broader corpus.
Two researchers independently categorized these streamers by their perceived racial category in broad buckets: white, person of color (PoC), and unavailable. Two streamers in the hate-raided sample were unable to be categorized due to the streamer either not including a video feed of their face or using a racially ambiguous virtual avatar (``VTuber'' model). We found that the majority of streamers were white in both the hate raided sample and the random sample. We found that the mainstream sample consisted of 41\% PoC streamers, which is higher than what we observed among hate raid victims (35\%). We note that there is a very large discrepancy between the proportion of PoC streamers identified through manual coding versus the Black/African American tags. This may be because these tags are not assigned by default, and in order to apply them, streamers must explicitly select them.
We performed a two-sample proportion test to evaluate whether the racial identities of the populations of (1) the victims of hate raids in our mainstream corpus and (2) our mainstream corpus differ.
In this case, we failed to reject the null hypothesis ($p>0.05$), meaning that we do not have evidence to say that the set of hate raids we quantified, which often missed the mark in the identities targeted in their content and the identities of the victim, disproportionately targeted PoC streamers as coded in this way. It is possible that this is in part due to an intersection of race and gender/sexual identity where the proportions of LGBTQ+-identifying streamers were unequal between racial groups, but we do not have the sample size to adequately test this within our sample. However, the above analysis does present evidence that hate raids occurred in different proportions across different identity tags, suggesting that tags may have been used as a targeting mechanism. 

Our sample size of 57 hate raided channels is inadequate to perform rigorous statistical testing to determine whether the broader set of attacks disproportionately targeted Black streamers, but we note that the proportion of anti-Black content in hate raid messages (98.3\%) is far greater than the proportion of Black streamers that we detected experiencing hate raids (10.5\%); further, only 2 of these Black streamers received hate raid messages that specifically contained racist anti-Black language, though implicitly racist and/or antisemitic undertones and references were still present in some. This disparity between the identities of the streamers and the kinds of hate spewed in their chats indicates that many of these attacks were indiscriminate in their targets---in most but not all cases, they were not tailored to the specific streamer, but rather contained a consistent breadth of hate regardless of their target. Paired with our tag analysis, we find that the extent of targeting in the observed hate raids may have relied on the usage of tags due to the ease of automation. Through this large-scale, quantitative perspective, we find evidence of another mode of hate raids that included identity-based attacks but did not align the content of their messages with the targeted streamers' identities. Rather, we see recurring themes and shared message text across hate raids in different channels regardless of the streamers' racial identities; we describe such general, reused hateful content sent en masse as ``canned hate.''

\subsubsection{Perspectives of Streamers from Targeted Communities}
While the \textit{content} of the hate raid messages primarily targeted Black and Jewish identities, analysis of tags revealed that \textit{streamers} who were attacked were disproportionately likely to be those using Black and/or LGBTQ+ tags. To better understand these nuances, we consider the perspectives of streamers from these targeted communities. We conducted a series of interviews with seven Twitch streamers (labeled as TS in quote attributions) that identified as Black and/or LGBTQ+. In this section, we discuss their accounts of how members of these communities perceived the targeted nature of hate raids and the different channels through which they were executed.

\paragraph{Degree of Targeting}
In our interviews with streamers, we found that streamers' experiences largely aligned with media descriptions of hate raids as a highly-targeted attack, often specifically targeted toward Black and LGBTQ+ creators on Twitch. Six of the seven streamers we interviewed explicitly described the primary targets of hate raids as Black, BIPOC, transgender, or LGBTQ+ communities. In addition to these commonly targeted demographics, two streamers noted that visibility also played a role in attackers' choice in targeted channels:
\begin{quote}
    \textit{``Specifically like one of my friends who has a bigger viewership, they've been affected a lot more.''} -- TS01
\end{quote}
\begin{quote}
    \textit{``[My experience] was very mild in comparison to other streamers' who were either vocally and proudly trans or Black or both, as those were absolutely the target demographics.''} -- TS06
\end{quote}
From these streamers' perspectives, viewership and reputation factored into which streamers were more likely to be targeted, in addition to their race and gender.

Additionally, while interviewees acknowledged that Black and LGBTQ+ communities have always been at-risk for hate and harassment on online platforms, three of seven participants stated that this wave of hate raids was drastically more severe than the attacks they had experienced before. For instance, one streamer described the hate raid they experienced in 2021 as ``arguably the worst raid'' and ``most egregious iteration'' they have seen to date (TS01). Per these accounts, we sought to examine what aspects of hate raids in 2021 distinguished them from previous attacks. Several streamers explained that this sharp peak in severity manifested in the persistence and scale of the attacks. The reported frequency of hate raids varied across the streamers; while one streamer stated that they were hate raided only once, two others observed a drastic increase in the duration and frequency of the hate raids they experienced firsthand and witnessed in other channels. One streamer recounted how they were hate raided for two weeks straight: 
\begin{quote}
    \textit{``The highlight was the first stream that they hit me in, I had a four and a half hour stream. They were in my stream for about three and a half of those hours, nonstop hate raiding me.''} -- TS01
\end{quote}
Another streamer (TS03) contrasted their experience with hate raids before and during this particularly active period, where before, hate-driven attacks occurred as a single burst that ``wasn't an all day, every day or an hours long thing'' and would ``die out for a while.'' However, in 2021, they witnessed hate raids that were far worse:
\begin{quote}
    \textit{``They were raided for three hours straight, just three hours of just following and trying to put messages in chat, trying dox them, whether it was by putting an address in a message or making a username with the address and just following incessantly.''} -- TS03
\end{quote}
While we did not find raids of this type within our dataset, we did not capture data from every targeted channel for reasons discussed above.

TS05 notes that the degree of automation played a key role in the impact of the threat; the usage of bots grew over time and later reached unprecedented scale---they would use a tool to block suspected bot accounts all night long, blocking 300,000 to 400,000 bots at a time. They described the churn of newly created and weaponized bot accounts as ``incessant and overwhelming.'' In addition, TS05 commented on the sharp growth in attacks throughout the summer:
\begin{quote}
    \textit{``It went from 0-100 in no time at all. But it got scary because they were finding personal information about me and throwing it into however many public internet locations as possible. I had 70+ people sending me screenshots of an address associated with me for weeks.''} -- TS05
\end{quote}
Both TS03 and TS05's experiences of these raids raised another concern---the targeted nature of the content of these messages. The carefully-crafted contents of these messages, in addition to expressing identity attacks against their targets, sought to threaten even the physical well-being of their targets via doxxing. The impact of such targeted attacks and the violent threats underlying doxxing even pushed one streamer to escalate their mitigation strategies beyond their stream:
\begin{quote}
    \textit{``Law enforcement got involved, I had to find a lawyer, [the attackers] were threatening violence against my children... It was a scary time for me.''} -- TS05
\end{quote}
TS05's experience was not unique. TS01 also expressed that others also experienced swatting as a result of being doxxed. These experiences of online harassment have manifested in potential psychological, physical, and even financial harm for already marginalized groups.

Through both the incessant and bespoke nature of these attacks, we found that these streamers' perceptions and experiences of hate raids defined them as highly-motivated attacks on individuals based on their identities, targeting Black and LGBTQ+ communities in particular; while attacks on these marginalized communities have always existed in online spaces, the severity and persistence of hate raids distinguishes them from what many members of these communities had experienced before.

\paragraph{Cross-Platform Attacks}
As explained by several streamers, the targeted nature of these attacks resulted in a varied set of vectors threatening their psychological and physical safety. To better define the range of threats hate raids posed to streamers, we asked each participant to describe their experiences with hate raids and what attack vectors were used. We find that four of the seven streamers envisioned hate raids on Twitch as one piece of a larger campaign of harassment, highlighting the multi-platform nature of these orchestrated attacks. For instance, TS01's address and phone number were released in public locations off Twitch, and attackers even made videos on other platforms to help disseminate their personal information. This was then leveraged to flood their phone with calls. Similarly, TS02 and TS04 noted that Discord was another platform of concern; Discord servers of targeted streamers were attacked, and some of the hate raids were organized in Discord servers. TS04 described the complexity of the multi-platform nature of these attacks:
\begin{quote}
    \textit{``Where I find that companies really fall flat is understanding the impact of things that happen on their platform, the things that are planned on their platform and committed on another one... I think this is part of the issue with some of these hate raids is that it is personal info being hit. It's people's personal stuff outside of hate raids, outside of Twitch being shared. It is also being called slurs in chat, and that's harmful absolutely to be called slurs in chat and stuff like that. But it's also the fear of, well, my full name just got shared or my address just got shared. For me, it was like my Discord got hit, which is a whole other platform.''} -- TS04
\end{quote}
TS05 echoes these concerns, acknowledging that while hate raids originated with Twitch, ``when someone makes it their mission to harm you, they'll look for whatever they can to access you.'' As a result, the high motivation involved in these attacks has raised questions and frustration within the community regarding platform accountability.

In tandem, our quantitative and qualitative data on hate raids indicate that the experiences of the streamers from Black and LGBTQ+ communities align with the media's portrayal of hate raids---that is, as highly-targeted and motivated attacks. However, through our quantitative analysis (Section~\ref{sss:quantitative}), we also identified a variation of hate raids that deviated from this depiction, a form of hate raids akin to subcultural trolling that did not target specific streamers according to their identities, instead using ``canned hate'' to spread hate against Black and Jewish identities en masse in popular channels with high visibility. That is, these hate raids contained identity-based attacks, but were spread across the platform indiscriminately, indicating that an eagerness to cause widely-visible, attention-grabbing chaos may have also motivated the attackers.
\subsection{Community response to hate raids}
\label{sec:response}
As hate raids swept the platform, the community's need and urgency for tools and resources to mitigate the threat grew. We performed a series of interviews with both streamers and bot developers involved with marginalized communities on Twitch to understand the following: (1) how streamers and their communities addressed the threat of hate raids in the short term, (2) what array of tools and resources were assembled to mitigate the impact of hate raids, (3) the efficacy of the grassroots organization for \#TwitchDoBetter and \#ADayOffTwitch, and (4) the longer-term effects of hate raids on streamers and their communities. 

\subsubsection{Short term responses by streamers}
Four of the seven streamers we interviewed expressed that they employed both proactive and reactive mitigation techniques to protect themselves from hate raids in the short term. The kinds of techniques varied, often depending on the severity of the threat. On one end of the spectrum, one streamer explained that because their attackers had escalated to threatening violence against their children, they involved law enforcement and retained a lawyer. While these vectors of attack were impossible to address solely on-platform, the majority of streamers experienced attacks that manifested within the Twitch ecosystem of chat and engagement notifications (e.g., follows and raids). As such, these streamers were able to mitigate some of the impact of hate raids via modifications to their streams' moderation protocols. One streamer added more users to their moderation teams, recruiting them from longtime members of their community who were ``constantly hanging out inside of the chat'' and ``offering up their services... so that they can keep an eye on the chat,'' a pattern previously identified in~\cite{seering2019moderator} and~\cite{seering2022whomoderates}. Several streamers detailed variations of informational-support seeking, resource aggregation, and development of new tools in ways similar to those previously detailed in crisis informatics literature. For instance, one streamer described Stream Deck presets that were helpful for an emergency response; a Stream Deck is a physical control pad with preset studio settings (e.g., switching media scenes, camera angles, executing chat commands). They described commands that they added for moderation purposes:
\begin{quote}
    \textit{``We added more commands to like basically put it in follower mode and to turn off the chat to where it's only emotes only so that they can't put in any hateful words. Shutting things down for 10 minutes, but with a push of a button.''} -- TS02
\end{quote}
Similarly, another streamer outlined channel lockdown protocols they followed for hate raids, incorporating the idea of a ``panic button'' into a human moderator pipeline to handle incidents post-facto: 
\begin{quote}
    \textit{``I had a panic button that turned off alerts, locked down chat, my mods would record times of incidents such as follow botting, we added different terms to the banned words list, had the highest auto mod settings available.''} -- TS05
\end{quote}

One variation of the Twitch Panic Button was developed and publicly advertised by \texttt{nutty}, a Twitch streamer, to be a rapid response mechanism integrated with a Stream Deck so that a single push of a button (or in customized cases, a voice-activated trigger phrase) enabled subscribers-only mode and cleared existing chat from both the chat client and the stream display~\cite{dacanay2021NEWSpanicbutton}. Furthermore, after performing damage control, \texttt{nutty}'s tool attempted to reclaim the stream space; for instance, in \texttt{nutty}'s stream, the button triggered changing background lights and snarky automatically generated messages. An official tool with functionality similar to a panic button, named ``Shield Mode,'' was rolled out by Twitch in late November, 2022.

In addition to the automated tools like the panic button, we found that streamers were aware of bot developers that developed bespoke features or new bots altogether to help handle the wave of hate raids. In our interviews, a streamer mentioned one bot in particular, Sery\_Bot:
\begin{quote}
    \textit{``Also there's an additional thing that has been added to a lot of... streamers chats called Sery\_Bot. Someone who isn't working for Twitch created a bot where it kind of shuts down all the other bots. Like it blocks them from being able to say anything. Or once they can come into your chat, it blocks them out. So a lot of us have added that.''} -- TS02
\end{quote}

Sery\_Bot was developed by Sery, a developer who also sometimes streamed on Twitch. On August 14, 2021, Sery publicly solicited the Twitch community via Twitter for examples of hate raid messages and other relevant information to begin developing his bot. In the span of just a couple of weeks, Sery developed a variety of features---for instance, text-based commands in IRC like \texttt{!hateraidon} that performed a similar function to the panic button. In addition to providing utility already provided by other tools, Sery\_Bot also integrated community-based block lists of account usernames to automatically check new chat messages against. Subsequent months entailed list updates, more feature development (e.g., checking account age of chatters and profile picture scanning for repeat offensive images, like swastikas). With the rollout of so many features so rapidly, Sery\_Bot became viral, and in just two months, it amassed over 55,000 integrations over different Twitch channels.

TS02 highlighted both the effectiveness and widespread adoption of such a third-party bot amongst streamers in the context of hate raids; however, they also indicated that discomfort with or distrust of technology---particularly third-party bots---may have inhibited the adoption of tools and resources meant to mitigate such threats within the community.

\subsubsection{Resources from tool developers and other community members}
The short-term responses from streamers alluded to the availability of community-sourced tools and streamers' reliance on them to combat hate raids. Our interviews with both streamers and bot developers illuminated the various kinds of tools and resources the community created and what the development process was like in response to real-time threats. Streamers' perspectives gave insights into what kinds of tools were visible and widespread throughout the community. Four streamers explicitly mentioned the use of third-party bots for hate raid mitigation or prevention.
One streamer noted that in addition to guides for making the aforementioned panic buttons, they were well-informed of the various bots that were developed individually by different bot developers, all for the purpose of responding to hate raids:
\begin{quote}
    \textit{``There was Smash Bot that was created. Mix It Up Bot, Sery Bot, StopHateBot, WiseBot, time out bot. All of those things were kind of developed.''} -- TS03
\end{quote}
Another streamer contrasted the fast wave of tool development by the community members with the poor communication and delayed response from Twitch:
\begin{quote}
    \textit{``And yet, for some reason we have six queers in a trench coat who have somehow made all these tools in the span of three days for us to use that no, they don't eradicate the issue, but they definitely helped kind of mitigate it immensely. We had [user] who was making full master lists along with [user] of all the bots that were being created which... Twitch I think should reach out to them and get that master list and pay them for their work and say, these are all bots that are out doing things that are worthwhile or valuable.''} -- TS01
\end{quote}
TS01 underscored that these tools were developed by members of the communities targeted by hate raids in a rapid-response fashion that Twitch as a platform simply could not; furthermore, TS01 emphasized that these quickly-developed tools were also effective in mitigating specifically the bot-mediated harassment even just with fairly naive methods.

To supplement our understanding of moderation bots' roles in the hate raid ecosystem, we interviewed two bot developers (referred to as BD in quote attributions) to understand (1) their perspectives and experiences with the community's needs, (2) Twitch as a platform for development, and (3) technical challenges they encountered while developing features for hate raids. One bot developer noted a sort of cat-and-mouse game between the community members and attackers, resulting in fast-paced changes in the sophistication of hate raids and applicable mitigation techniques. This bot developer remarked on the primitive, manual nature of how hate raids started, only to quickly become more coordinated and varied:
\begin{quote}
    \textit{``So they weren't that organized back then, started with a couple guys who just came into [a streamer]'s voice chat while he was streaming Phasmophobia, and they were shouting the N word, and then he gave them a really strong reaction by immediately ending the stream, deleting the VOD and so on. So they came back and spammed all nasty stuff in chat.''} -- BD02
\end{quote}
This bot developer also illuminated some of the motivation for hate raid participants: to elicit a strong, disruptive reaction from the streamer. Similarly, there were approaches that the larger community initially employed, such as joining the attackers' Discord servers where hate raids are organized, that the attackers responded to:
\begin{quote}
    \textit{``And it's also a little bit of a double-edged sword because we originally used to enter the offensive Discord servers where they would gather up and organize these hate raids with a second account, and then report the Discord server and the respective messages. And now of course, they learned that we do that, and they also set up a similar set of security measures.''} -- BD02
\end{quote}
As such, both the attackers and the defenders in the hate raid ecosystem were made aware of what strategies the other side was employing, and they adapted their methods in response. From one bot developer's perspective, they analyzed the threat of hate raid messages and identified that attackers used ``automated tools to just spam the chat,'' and in response began developing a bot as a counter measure. This bot developer noted that the hate spam sent by an army of bots was difficult to mitigate with manual moderation, so identified a bot-mediated moderation approach as an effective way to respond to the attack.

\subsubsection{Collective action for \#TwitchDoBetter/\#ADayOffTwitch}

In addition to the different ways the community attempted to better protect themselves from hate raids, there were also attempts by the community to raise awareness through collective action. Two hashtags, \#TwitchDoBetter and \#ADayOffTwitch, rallied support from Twitch users via Twitter. \#TwitchDoBetter was started in an attempt to raise awareness of the harassment of targeted creators on Twitch. Subsequently, \#ADayOffTwitch was a boycott of Twitch that took place on September 1, 2021, meaning participants would not stream, watch streams, or participate in any chats. This day took place with hopes that reduced engagement on the platform would highlight the urgency of better safety for creators on Twitch. We found through our interviews with streamers that 
 while the community was able to raise awareness through these movements, there was some disagreement about their long-term impact within the community. One streamer we interviewed who was a co-organizer of the walkout felt that the movements were successful in meeting what they perceived to be the goal, which was to raise awareness:
\begin{quote}
    \textit{``I think it worked in the way that I had hoped. We raised awareness. We actively called on a MAJOR streaming platform to make changes and we're seeing the fruits of our labor. It's not always about money like so many bigger streamers commented. Sometimes we have to understand that reputation is a currency.''} -- TS05
\end{quote}
However, setting an open-ended goal of raising awareness for these community-organized movements did not satisfy another participant. In contrast, TS01 felt that ultimately, these movements failed to narrow in on concrete demands of Twitch and therefore failed to be as effective and impactful as they could have been:
\begin{quote}
    \textit{``I feel like they should have done a better job of sitting down and figuring out an actualized list of demands. Why are we taking a day off of Twitch? Because at face value, the reason we're taking the day off of Twitch is because hate raids suck and that's a true assessment. But what about after that? Why do those hate raids suck? What do we want to see to address those hate raids? How do we want Twitch to address it? How do we want Twitch to actually get engaged more about it? How do we want Twitch to respond to this? How does that carry over into future endeavors? What does that look like for a conversation around how they need to update their security? There was just little to no demand to be had whatsoever. And so, what could have been an actual movement or an actual kind of protest type of thing, just wound up being plainly speaking, a bunch of people just not logging in.''} -- TS01
\end{quote}
While these movements were organized within the community, perceptions of their goals and effectiveness varied throughout. Still, despite conflict around the goal and organization of the movement, \#ADayOffTwitch did indeed significantly impact the number of viewers and streamers engaging with the platform; per one external estimate, this movement led to up to 15\% less engagement on Twitch overall during the walkout~\cite{parrish2021NEWSprotests}.

\subsubsection{Longer-term impacts of hate raids on streamers}
Even with all of the mitigation attempts and movements to raise awareness, hate raids undoubtedly caused distress and skepticism throughout the community. Our interviews with streamers indicate that the visceral nature of these attacks paired with Twitch's response has largely shaped their views of the platform as unprepared and detached from the community's suffering. All seven of the participants expressed disappointment with Twitch's failure to consider abuse protections proactively, the slow rollout of features and tools to mitigate the harm of hate raids, and poor communication with between Twitch and stakeholders. One streamer expressed resigned frustration that this experience had been consistent with Twitch's attitudes toward protecting its at-risk communities in the past:
\begin{quote}
    \textit{``I think Twitch's response has been absolutely abysmal. I think that very frankly speaking, it's pretty pathetic. Twitch has a longstanding historical track record of not knowing how to communicate ever at all. So, while I do think that their communication for this was abysmal, I would be remiss to omit the part where it is exactly what I expected them to do. And Twitch is going to continue falling on their face over topics of this nature and conversations of this type every single time so long as they insist that silence is the best solution. And they need to do better than put out a simple tweet saying, `We hear you, we see you and we want you to know that we care, we promise.'''} -- TS01
\end{quote}
Streamers also felt that poor communication even around existing mitigation tools led to unnecessary chaos during hate raids. One streamer noted that enabling two-factor authentication was a common suggestion by the community and Twitch for streamers to protect themselves. However, these suggestions conflate the verification of user identity with that of accounts chatting in that user's channel. Therefore, even if there are tools that may better protect individuals, poor communication may lead to the misuse of the tool or misinterpretation of the protections it actually offers.

In addition to disappointment with Twitch's communication response, several streamers lamented the lack of tooling Twitch had prepared for such attacks, even for features that had been requested in the past or existed on other platforms:
\begin{quote}
    \textit{``The chat verification tools [released at the tail end of the hate raids] are really nice, and I think that that's what a lot of people have wanted for so long. I'm not sure exactly why it took them that long to implement. I feel like it should have been implemented.''} -- TS03
\end{quote}
\begin{quote}
    \textit{``I went to Twitch HQ for [a Black History summit in the past], and that was one of the things that all of us echoed and said and was like, `If I banned someone, they should not be able to continue consuming my content.' It needs to be like some of these other sites. Twitter and Facebook are perfect examples. When I block somebody, it's scorched earth. As far as they're concerned, I no longer exist to them. That's what it needs to be.''} -- TS01
\end{quote}
These embody some of the frustrations that streamers have had for baseline protection features that the community had been wanting for years. The overall emotional harm of hate raids was even enough to dissuade some members to leave Twitch or even streaming altogether; one streamer explains that the hate raids gave them a lot of anxiety, and that they know streamers who ``walked away entirely because of that anxiety and distress.'' Another streamer expressed concern that their protective measures to prevent anomalous viewership might even affect their stream's long-term growth:
\begin{quote}
    \textit{``We have things that we're trying to do at all times and if we blockade the people who want to watch us, they are going to inherently want to move on elsewhere.''} -- TS01
\end{quote}
More broadly, the threat of hate raids and their impact on streamers in the future still looms over several of the participants. One streamer noted that, while there may be a sense of fatigue among the community concerning hate raids, the lack of recent publicity over hate raids may not be an indication that the larger threat has passed:
\begin{quote}
    \textit{``I really think that it's either happen[ing] less or because we've been dealing with it now for so long, people are just... There's only so many times that you can post and be like, `Yep, got hate raided again today. Yep, got hate raided again today.' So that could also be a factor as to why I'm not seeing it as much on Twitter.''} -- TS03
\end{quote}

\section{Discussion}
In this paper, we make three primary contributions: (1) the descriptive characterization of a novel form of long-term harassment campaigns on livestreaming platforms; (2) the definition of hate raids as a dually-motivated phenomenon: first as a hate-driven attack, and second as an act of seeking attention; (3) the observation that members of targeted communities rapidly responded to the threat of hate raids to address the shortcomings of protections provided by Twitch. In each of the following paragraphs, we elaborate on these contributions.

\subsection{Characterization of Hate Raids}
Although hate raids on Twitch caused significant disruption and emotional harm to streamers, these attacks were relatively technically unsophisticated. Accounts were created en masse (likely in an automated fashion) to serve a single purpose, hateful comments were largely identical across channels, and user-specified identity tags were operationalized to attack marginalized groups. Many of these tactics might have been prevented if Twitch had followed established trust and safety practices like rate-limiting account creation~\cite{thomas2011suspended}, adding a delay between account creation and platform participation, deploying additional identity verification requirements (e.g., SMS or phone)~\cite{thomas2014dialing},\footnote{Phone verification was added as a feature during the later phases of hate raids, indicating that it may have been under development but not yet released when this wave of hate raids began.} and protecting at-risk streamers by safeguarding automated access to sensitive data, such as identity-based channel tags. 

Although there are trade-offs between adding friction in joining communities and protecting users from abuse, the security practices employed by Twitch at the time of this wave of hate raids did not deter these relatively unsophisticated attacks. The broader history of Trust and Safety is often characterized by \textit{reactive} feature development as attack vectors become apparent on each specific platform, but a common set of forms of attacks have appeared many times throughout the history of social platforms and, as in this case, they do not become substantially more sophisticated as they are ported from platform to platform~\cite{gordon1994irc,mariconti2019you,massanari2017gamergate}. The appearance of these attacks and the form that they take on any new platform is often predictable, and it is much easier to build safeguards during earlier development phases than to be forced to reactively add them under time pressure when crises arise. Future community-driven platforms should prioritize the allocation of resources to teams developing defensive tactics a necessary first step for curbing online abuse of this nature before it causes significant harm. 

The qualitative accounts of streamers' experiences that we examine affirm that highly-targeted hate raids can lead to long-term emotional distress and can even threaten streamers' physical safety. While Twitch has begun to take steps to combat hate raids via automated tooling (e.g., AutoMod), optional account verification methods, and the aforementioned Shield Mode, the threat of highly-motivated hate raids coordinated off-platform continues to loom over its streamers. In March 2022, a wave of hate raids orchestrated by streamers on Cozy.tv, a livestreaming platform founded by far-right white nationalist Nick Fuentes, hit Twitch, this time targeting women and LGBTQ+ streamers with homophobic, transphobic, and misogynistic messages in their Twitch channels, direct messages, and Discord servers~\cite{Polhamus2022NEWShateraids}.\footnote{These hate raids were performed manually, and as such would likely not have been deterred by security measures designed to prevent automated attacks from bots; however, their occurrence represents a continued threat to streamers from marginalized groups on the platform.} As hate raids continue to threaten streamers with varying degrees of off-platform coordination, legitimate user participation, and bot account manipulation, the need for platforms to consider both increasingly sophisticated threat models and historically common patterns of attacks has only grown. Platforms must consider preemptively what their policies, protection, and communication processes will be, and by designing these mechanisms for the needs of their at-risk communities, they can better protect all of their users. The design of proactive prevention measures that do not disproportionately burden or disadvantage marginalized communities---with respect to their online engagement and technical overhead---remains an important question for future research and development.

\subsection{Dual Motivation of Hate Raids}
We draw several primary characteristics from Marwick \cite{marwick2021networked} and Phillips \cite{phillips2015we} as a baseline to compare hate raids with: first, per Phillips, subcultural trolling benefits from (and to some extent relies on) amplification \cite[pp.~3--6, 56--61]{phillips2015we}, and fits within existing media narratives, often referencing mainstream concepts and/or publicized events \cite[pp.~115--118]{phillips2015we} in absurd or repurposed ways. Second, per Marwick, morally-motivated networked harassment also benefits from amplification, but it also relies heavily on identity and identity conflicts to justify harassment campaigns that have none of the underlying absurd logic that characterize subcultural trolling \cite[pp.~5--8]{marwick2021networked}. Moreover, where subcultural trolling originates from specific communities, often in planned, targeted attacks, morally-motivated networked harassment often originates more organically and is partially self-amplifying through the properties of networks such as those on Twitter. As we discussed in Section~\ref{sec:characterization}, the hate raids on Twitch share variants of each of these characteristics, and we therefore argue that they lie in a space between subcultural trolling and morally-motivated networked harassment.

\subsection{Stakeholder Rapid Response}

We observed many similar behaviors in Twitch hate raids that occur during natural disaster response as documented in crisis informatics literature --- informational-support seeking, aggregation of resources, and development of new tools and technologies to address specialized needs arising from the crisis. We also observed the use of social media for social support-seeking and solidarity, even leading to the organization of a significant protest.

During the hate raids, there was no formal organization (e.g., a state or federal government) coordinating public response, as is often the case in the aftermath of natural disasters; while Twitch did respond to the hate raids in several ways, these did not involve coordinating with its users at any scale. As such, responses to hate raids more closely resembled those documented in literature on longer-term conflicts where public institutions play less of a role because they have been weakened as a result of the conflict \cite[pp.~2--3]{monroyhernandez2013correspondents}. 

Users were able to respond to rapidly evolving situations during the hate raids in ways that brought relief to their communities far more quickly than Twitch was able to. They developed and rapidly iterated on tools to counter the attacks and improved those tools as attacks changed. The first of these tools appeared within days of when the hate raids started to gain public attention. Users also created guides on how to use Twitch's moderation features and Discord's moderation features (for streamers who had servers affiliated with their streams), and on how streamers could better protect their personal information. Guides for all of these already existed in forms created by the respective platforms, but the community-created guides gained significantly more traction in this case because of their applicability to the specific circumstances of hate raids and because of the shared trust between community members.

With this work we do not mean to suggest that, because users were effective in rapidly responding to these issues, Twitch should cede their authority to users on Trusty \& Safety issues. Instead, we note that Twitch and its users each have different strengths in how they are able to respond, and that Twitch and other platforms with similar moderation structures could gain much value from better communication and collaboration with users on moderation problems that arise. Volunteer moderators' domain-specific knowledge and reputational trust paired with the findings from prior work showing that experienced moderators can successfully onboard volunteers into new moderation contexts~\cite{seering2022pride} suggests that Twitch as a platform can gain insight and trust from their users by building connections with power users (e.g., prominent community tool developers like Sery). By consulting with such users, Twitch can also improve the dissemination of resource guides and the visibility of community-built tools. This access to information may be particularly effective in enhancing coordinated action because users are far more agile than the platform in organizing and producing tools to respond to imminent threats. As both Seering et al. and Roberts suggest~\cite{seering2022pride,roberts2016commercial}, the use of volunteer moderation for commercial platforms brings to question the ethics in the division of labor between volunteers and platforms. We argue that platforms should consult with power users to improve communication and tooling, and that these platforms should consider paying such power users for their valuable, contextual knowledge to compensate them for their large contributions to their communities.

Finally, we reiterate the recommendations of prior work~\cite{Blackwell2017classification} rooted in intersectional feminist theory: that platforms must center the needs of their most marginalized, vulnerable users in their design. Platforms designed around existing structural inequalities recreate and further disseminate these systems of oppression~\cite{Noble2018algorithms}. We argue that addressing the needs of the oppressed more effectively encompasses the needs of all users, allowing platforms to be better prepared to mitigate inevitable attempts of abuse.




\subsection{Limitations}
We acknowledge that our analysis is not based on a comprehensive view of the platform. Because smaller communities may have a tacit expectation of privacy, we intentionally did not collect chat data from channels with less than an average of 100 viewers. However, many communities targeted by hate raids were not necessarily large, mainstream channels. According to Twitch, as of 2018, 81.5\% of its creators and viewers were male~\cite{yosilewitz2021NEWSstats}, and user surveys have shown that a majority are white. As such, we expect that some of the highly-targeted hate raiding behavior was not captured in our large-scale data collection methodology. Furthermore, our hate raid detection mechanism was based on community-aggregated lists of known malicious bots. Because of this, we may not have detected categories of hate raids that were not actively documented by community members. This likely narrows the variance in attack structure and message content flagged in our dataset. Even with these limitations, however, we argue that our quantitative perspective still provides insights into various technical characteristics and attacker motivations of hate raids. Particularly, when paired with our qualitative results that \textit{specifically} seek the perspectives of targeted community members, we believe that we are able to capture multiple facets of a nuanced and dynamic threat model.


\section{Conclusion}
Our large-scale quantitative measurement of hate raids across mainstream channels on Twitch and interviews with community members from targeted groups confirm that hate raids are indeed highly-targeted and hate-driven attacks. Our quantitative analysis reveals an additional mode of hate raid, however, that is similar to subcultural trolling and networked harassment. We find that the technical characteristics of these attacks mirror many of the naïve methods of other forms of online abuse, such as spam. The content of these hate raid messages are deeply entrenched in two main hateful ideologies: anti-Black racism and
antisemitism. Our interviews demonstrate the various approaches---both proactive and reactive---to defense that the community took in response to hate raids. Our analysis furthers our understanding of the complexities in the ecosystem surrounding hate raids, highlights lessons to be learned in designing proactive harassment mitigation into a platform from the start, and brings attention to the interplay between platform and community governance in the face of a collective crisis.
\section{Acknowledgements}
This work was supported in part by grants \#2030859 and \#2127309 to the Computing Research Association for the CIFellows Project and
NSF Graduate Research Fellowship \#DGE-1656518.
We would like to thank the participants in this study for their time and openness in discussing their experiences, as well as Sery and PleasantlyTwstd for expert feedback on the nature of hate raids during the study. We would also like to thank Michael Bernstein for feedback on study design and communication of findings.

\bibliographystyle{ACM-Reference-Format}
\bibliography{paper}


\begin{thebibliography}{62}


\ifx \showCODEN    \undefined \def \showCODEN     #1{\unskip}     \fi
\ifx \showDOI      \undefined \def \showDOI       #1{#1}\fi
\ifx \showISBNx    \undefined \def \showISBNx     #1{\unskip}     \fi
\ifx \showISBNxiii \undefined \def \showISBNxiii  #1{\unskip}     \fi
\ifx \showISSN     \undefined \def \showISSN      #1{\unskip}     \fi
\ifx \showLCCN     \undefined \def \showLCCN      #1{\unskip}     \fi
\ifx \shownote     \undefined \def \shownote      #1{#1}          \fi
\ifx \showarticletitle \undefined \def \showarticletitle #1{#1}   \fi
\ifx \showURL      \undefined \def \showURL       {\relax}        \fi
\providecommand\bibfield[2]{#2}
\providecommand\bibinfo[2]{#2}
\providecommand\natexlab[1]{#1}
\providecommand\showeprint[2][]{arXiv:#2}

\bibitem[\protect\citeauthoryear{Blackwell, Dimond, Schoenebeck, and
  Lampe}{Blackwell et~al\mbox{.}}{2017}]%
        {Blackwell2017classification}
\bibfield{author}{\bibinfo{person}{Lindsay Blackwell}, \bibinfo{person}{Jill
  Dimond}, \bibinfo{person}{Sarita Schoenebeck}, {and} \bibinfo{person}{Cliff
  Lampe}.} \bibinfo{year}{2017}\natexlab{}.
\newblock \showarticletitle{Classification and Its Consequences for Online
  Harassment: Design Insights from HeartMob}.
\newblock \bibinfo{journal}{\emph{Proc. ACM Hum.-Comput. Interact.}}
  \bibinfo{volume}{1}, \bibinfo{number}{CSCW}, Article \bibinfo{articleno}{24}
  (\bibinfo{date}{Dec.} \bibinfo{year}{2017}), \bibinfo{numpages}{19}~pages.
\newblock
\showISSN{2573-0142}
\urldef\tempurl%
\url{https://doi.org/10.1145/3134659}
\showDOI{\tempurl}


\bibitem[\protect\citeauthoryear{Cai and Wohn}{Cai and Wohn}{2022}]%
        {cai2022coordination}
\bibfield{author}{\bibinfo{person}{Jie Cai} {and}
  \bibinfo{person}{Donghee~Yvette Wohn}.} \bibinfo{year}{2022}\natexlab{}.
\newblock \showarticletitle{Coordination and Collaboration: How do Volunteer
  Moderators Work as a Team in Live Streaming Communities?}
\newblock \bibinfo{journal}{\emph{Proceedings of the 2019 CHI Conference on
  Human Factors in Computing Systems}} (\bibinfo{year}{2022}).
\newblock


\bibitem[\protect\citeauthoryear{Castillo}{Castillo}{2016}]%
        {castillo2016big}
\bibfield{author}{\bibinfo{person}{Carlos Castillo}.}
  \bibinfo{year}{2016}\natexlab{}.
\newblock \bibinfo{booktitle}{\emph{Big crisis data: social media in disasters
  and time-critical situations}}.
\newblock \bibinfo{publisher}{Cambridge University Press},
  \bibinfo{address}{Cambridge, UK}.
\newblock


\bibitem[\protect\citeauthoryear{Chandrasekharan, Gandhi, Mustelier, and
  Gilbert}{Chandrasekharan et~al\mbox{.}}{2019}]%
        {chandrasekharan2019crossmod}
\bibfield{author}{\bibinfo{person}{Eshwar Chandrasekharan},
  \bibinfo{person}{Chaitrali Gandhi}, \bibinfo{person}{Matthew~Wortley
  Mustelier}, {and} \bibinfo{person}{Eric Gilbert}.}
  \bibinfo{year}{2019}\natexlab{}.
\newblock \showarticletitle{Crossmod: A Cross-Community Learning-Based System
  to Assist Reddit Moderators}.
\newblock \bibinfo{journal}{\emph{Proc. ACM Hum.-Comput. Interact.}}
  \bibinfo{volume}{3}, \bibinfo{number}{CSCW}, Article \bibinfo{articleno}{174}
  (\bibinfo{date}{Nov.} \bibinfo{year}{2019}), \bibinfo{numpages}{30}~pages.
\newblock
\urldef\tempurl%
\url{https://doi.org/10.1145/3359276}
\showDOI{\tempurl}


\bibitem[\protect\citeauthoryear{Cheng, Bernstein, Danescu-Niculescu-Mizil, and
  Leskovec}{Cheng et~al\mbox{.}}{2017}]%
        {cheng2017anyone}
\bibfield{author}{\bibinfo{person}{Justin Cheng}, \bibinfo{person}{Michael
  Bernstein}, \bibinfo{person}{Cristian Danescu-Niculescu-Mizil}, {and}
  \bibinfo{person}{Jure Leskovec}.} \bibinfo{year}{2017}\natexlab{}.
\newblock \showarticletitle{Anyone Can Become a Troll: Causes of Trolling
  Behavior in Online Discussions}. In \bibinfo{booktitle}{\emph{Proceedings of
  the 2017 ACM Conference on Computer Supported Cooperative Work and Social
  Computing}} (Portland, Oregon, USA) \emph{(\bibinfo{series}{CSCW '17})}.
  \bibinfo{publisher}{ACM}, \bibinfo{address}{New York, NY, USA},
  \bibinfo{pages}{1217--1230}.
\newblock
\showISBNx{978-1-4503-4335-0}
\urldef\tempurl%
\url{https://doi.org/10.1145/2998181.2998213}
\showDOI{\tempurl}


\bibitem[\protect\citeauthoryear{Cheong and Lee}{Cheong and Lee}{2011}]%
        {cheong2011terrorism}
\bibfield{author}{\bibinfo{person}{Marc Cheong} {and} \bibinfo{person}{Vincent
  C~S Lee}.} \bibinfo{year}{2011}\natexlab{}.
\newblock \showarticletitle{{A microblogging-based approach to terrorism
  informatics: Exploration and chronicling civilian sentiment and response to
  terrorism events via Twitter}}.
\newblock \bibinfo{journal}{\emph{Information Systems Frontiers}}
  \bibinfo{volume}{13}, \bibinfo{number}{1} (\bibinfo{year}{2011}),
  \bibinfo{pages}{45--59}.
\newblock
\showISSN{1572-9419}
\urldef\tempurl%
\url{https://doi.org/10.1007/s10796-010-9273-x}
\showDOI{\tempurl}


\bibitem[\protect\citeauthoryear{Cooney}{Cooney}{[n.d.]}]%
        {cooney2021NEWSviewers}
\bibfield{author}{\bibinfo{person}{Bill Cooney}.}
  \bibinfo{year}{[n.d.]}\natexlab{}.
\newblock \showarticletitle{New Twitch stats reveal how few viewers are needed
  to be a “top” streamer}.
\newblock \bibinfo{journal}{\emph{Dexerto}} (\bibinfo{year}{[n.\,d.]}).
\newblock
\urldef\tempurl%
\url{https://www.dexerto.com/entertainment/new-twitch-stats-reveal-how-few-viewers-are-needed-to-be-a-top-streamer-1527638/}
\showURL{%
\tempurl}


\bibitem[\protect\citeauthoryear{Cote}{Cote}{2017}]%
        {cote2017strategies}
\bibfield{author}{\bibinfo{person}{Amanda~C Cote}.}
  \bibinfo{year}{2017}\natexlab{}.
\newblock \showarticletitle{{“I Can Defend Myself”: Women's Strategies for
  Coping With Harassment While Gaming Online}}.
\newblock \bibinfo{journal}{\emph{Games and Culture}} \bibinfo{volume}{12},
  \bibinfo{number}{2} (\bibinfo{year}{2017}), \bibinfo{pages}{136--155}.
\newblock
\urldef\tempurl%
\url{https://doi.org/10.1177/1555412015587603}
\showDOI{\tempurl}


\bibitem[\protect\citeauthoryear{Creswell}{Creswell}{2013}]%
        {creswell2013qualitative}
\bibfield{author}{\bibinfo{person}{John~W Creswell}.}
  \bibinfo{year}{2013}\natexlab{}.
\newblock \bibinfo{booktitle}{\emph{Qualitative Inquiry and Research Design:
  Choosing Among Five Traditions}}.
\newblock \bibinfo{publisher}{SAGE}, \bibinfo{address}{Thousand Oaks, CA}.
\newblock


\bibitem[\protect\citeauthoryear{Dacanay}{Dacanay}{[n.d.]}]%
        {dacanay2021NEWSpanicbutton}
\bibfield{author}{\bibinfo{person}{Ralston Dacanay}.}
  \bibinfo{year}{[n.d.]}\natexlab{}.
\newblock \showarticletitle{Twitch Streamer Creates Third-Party 'Panic Button'
  to Counter Hate Raids}.
\newblock \bibinfo{journal}{\emph{DBLTAP}} (\bibinfo{year}{[n.\,d.]}).
\newblock
\urldef\tempurl%
\url{https://www.dbltap.com/posts/twitch-streamer-creates-third-party-panic-button-to-counter-hate-raids-01fekqmgacvw}
\showURL{%
\tempurl}


\bibitem[\protect\citeauthoryear{D'Anastasio}{D'Anastasio}{[n.d.]}]%
        {danastasio2021NEWShateraids}
\bibfield{author}{\bibinfo{person}{Cecilia D'Anastasio}.}
  \bibinfo{year}{[n.d.]}\natexlab{}.
\newblock \showarticletitle{Twitch Sues Users Over Alleged ‘Hate Raids’
  Against Streamers}.
\newblock \bibinfo{journal}{\emph{Wired}} (\bibinfo{year}{[n.\,d.]}).
\newblock
\urldef\tempurl%
\url{https://www.wired.com/story/twitch-sues-users-over-alleged-hate-raids/}
\showURL{%
\tempurl}


\bibitem[\protect\citeauthoryear{Datta and Adar}{Datta and Adar}{2019}]%
        {datta2019extracting}
\bibfield{author}{\bibinfo{person}{Srayan Datta} {and} \bibinfo{person}{Eytan
  Adar}.} \bibinfo{year}{2019}\natexlab{}.
\newblock \showarticletitle{Extracting inter-community conflicts in reddit}. In
  \bibinfo{booktitle}{\emph{Proceedings of the international AAAI conference on
  Web and Social Media}}.
\newblock


\bibitem[\protect\citeauthoryear{Dibbell}{Dibbell}{1993}]%
        {Dibbell1993rape}
\bibfield{author}{\bibinfo{person}{Julian Dibbell}.}
  \bibinfo{year}{1993}\natexlab{}.
\newblock \showarticletitle{A Rape in Cyberspace: How an Evil Clown, a Haitian
  Trickster Spirit, Two Wizards, and a Cast of Dozens Turned a Database Into a
  Society}.
\newblock \bibinfo{journal}{\emph{The Village Voice}}
  \bibinfo{volume}{December 23} (\bibinfo{year}{1993}),
  \bibinfo{pages}{36--42}.
\newblock
\urldef\tempurl%
\url{https://www.villagevoice.com/2005/10/18/a-rape-in-cyberspace/}
\showURL{%
\tempurl}


\bibitem[\protect\citeauthoryear{Elmas, Overdorf, {\"O}zkalay, and
  Aberer}{Elmas et~al\mbox{.}}{2022}]%
        {elmas2022astroturf}
\bibfield{author}{\bibinfo{person}{Tu\u{g}rulcan Elmas},
  \bibinfo{person}{Rebekah Overdorf}, \bibinfo{person}{Ahmed~Furkan
  {\"O}zkalay}, {and} \bibinfo{person}{Karl Aberer}.}
  \bibinfo{year}{2022}\natexlab{}.
\newblock \showarticletitle{Ephemeral Astroturfing Attacks: The Case of Fake
  Twitter Trends}. In \bibinfo{booktitle}{\emph{EuroS\&P '22}}.
\newblock


\bibitem[\protect\citeauthoryear{Fox and Tang}{Fox and Tang}{2017}]%
        {fox2017womens}
\bibfield{author}{\bibinfo{person}{Jesse Fox} {and} \bibinfo{person}{Wai~Yen
  Tang}.} \bibinfo{year}{2017}\natexlab{}.
\newblock \showarticletitle{{Women's experiences with general and sexual
  harassment in online video games: Rumination, organizational responsiveness,
  withdrawal, and coping strategies}}.
\newblock \bibinfo{journal}{\emph{New Media {\&} Society}}
  \bibinfo{volume}{19}, \bibinfo{number}{8} (\bibinfo{year}{2017}),
  \bibinfo{pages}{1290--1307}.
\newblock
\urldef\tempurl%
\url{https://doi.org/10.1177/1461444816635778}
\showDOI{\tempurl}


\bibitem[\protect\citeauthoryear{Gordon}{Gordon}{1994}]%
        {gordon1994irc}
\bibfield{author}{\bibinfo{person}{Sara Gordon}.}
  \bibinfo{year}{1994}\natexlab{}.
\newblock \showarticletitle{IRC and Security --- Can the two co-exist?}
\newblock  (\bibinfo{year}{1994}).
\newblock


\bibitem[\protect\citeauthoryear{Gray}{Gray}{2017}]%
        {gray2017urban}
\bibfield{author}{\bibinfo{person}{Kishonna~L Gray}.}
  \bibinfo{year}{2017}\natexlab{}.
\newblock \showarticletitle{They’re just too urban”: Black gamers streaming
  on Twitch}.
\newblock In \bibinfo{booktitle}{\emph{Digital Sociologies}},
  \bibfield{editor}{\bibinfo{person}{Jessie Daniels}, \bibinfo{person}{Karen
  Gregory}, {and} \bibinfo{person}{Tressie McMillan~Cottom}} (Eds.).
  Vol.~\bibinfo{volume}{1}. \bibinfo{publisher}{Policy Press},
  \bibinfo{address}{Bristol, England}, Chapter~22, \bibinfo{pages}{355--368}.
\newblock


\bibitem[\protect\citeauthoryear{Grayson}{Grayson}{[n.d.]a}]%
        {grayson2021NEWStwitchdobetter}
\bibfield{author}{\bibinfo{person}{Nathan Grayson}.}
  \bibinfo{year}{[n.d.]}\natexlab{a}.
\newblock \showarticletitle{Marginalized streamers beg Twitch to ‘do
  better’ in wake of hate raids, poor pay}.
\newblock \bibinfo{journal}{\emph{The Washington Post}}
  (\bibinfo{year}{[n.\,d.]}).
\newblock
\urldef\tempurl%
\url{https://www.washingtonpost.com/video-games/2021/08/11/twitch-do-better-hate-raids/}
\showURL{%
\tempurl}


\bibitem[\protect\citeauthoryear{Grayson}{Grayson}{[n.d.]b}]%
        {grayson2021NEWShateraids}
\bibfield{author}{\bibinfo{person}{Nathan Grayson}.}
  \bibinfo{year}{[n.d.]}\natexlab{b}.
\newblock \showarticletitle{Twitch hate raids are more than just a Twitch
  problem, and they’re only getting worse}.
\newblock \bibinfo{journal}{\emph{The Washington Post}}
  (\bibinfo{year}{[n.\,d.]}).
\newblock
\urldef\tempurl%
\url{https://www.washingtonpost.com/video-games/2021/08/25/twitch-hate-raids-streamers-discord-cybersecurity/}
\showURL{%
\tempurl}


\bibitem[\protect\citeauthoryear{Herring, Job-Sluder, Scheckler, and
  Barab}{Herring et~al\mbox{.}}{2002}]%
        {herring2002}
\bibfield{author}{\bibinfo{person}{Susan Herring}, \bibinfo{person}{Kirk
  Job-Sluder}, \bibinfo{person}{Rebecca Scheckler}, {and}
  \bibinfo{person}{Sasha Barab}.} \bibinfo{year}{2002}\natexlab{}.
\newblock \showarticletitle{{Searching for Safety Online: Managing ``Trolling''
  in a Feminist Forum}}.
\newblock \bibinfo{journal}{\emph{The Information Society}}
  \bibinfo{volume}{18}, \bibinfo{number}{5} (\bibinfo{year}{2002}),
  \bibinfo{pages}{371--384}.
\newblock
\urldef\tempurl%
\url{https://doi.org/10.1080/01972240290108186}
\showDOI{\tempurl}


\bibitem[\protect\citeauthoryear{Horetski}{Horetski}{[n.d.]}]%
        {horetski2022NEWShateraids}
\bibfield{author}{\bibinfo{person}{Dylan Horetski}.}
  \bibinfo{year}{[n.d.]}\natexlab{}.
\newblock \showarticletitle{Twitch hate raids return in massive wave of attacks
  on LGBTQIA+ streamers}.
\newblock \bibinfo{journal}{\emph{Dexerto}} (\bibinfo{year}{[n.\,d.]}).
\newblock
\urldef\tempurl%
\url{https://www.dexerto.com/entertainment/twitch-hate-raids-return-in-massive-wave-of-attacks-on-lgbtqia-streamers-1781574/}
\showURL{%
\tempurl}


\bibitem[\protect\citeauthoryear{Jiang, Kiene, Middler, Brubaker, and
  Fiesler}{Jiang et~al\mbox{.}}{2019}]%
        {Jiang2019voice}
\bibfield{author}{\bibinfo{person}{Jialun~Aaron Jiang},
  \bibinfo{person}{Charles Kiene}, \bibinfo{person}{Skyler Middler},
  \bibinfo{person}{Jed~R. Brubaker}, {and} \bibinfo{person}{Casey Fiesler}.}
  \bibinfo{year}{2019}\natexlab{}.
\newblock \showarticletitle{Moderation Challenges in Voice-based Online
  Communities on Discord}.
\newblock \bibinfo{journal}{\emph{Proc. ACM Hum.-Comput. Interact.}}
  \bibinfo{volume}{3}, \bibinfo{number}{CSCW}, Article \bibinfo{articleno}{55}
  (\bibinfo{date}{Nov.} \bibinfo{year}{2019}), \bibinfo{numpages}{23}~pages.
\newblock
\showISSN{2573-0142}
\urldef\tempurl%
\url{https://doi.org/10.1145/3359157}
\showDOI{\tempurl}


\bibitem[\protect\citeauthoryear{Kanich, Kreibich, Levchenko, Enright, Voelker,
  Paxson, and Savage}{Kanich et~al\mbox{.}}{2008}]%
        {kanich2008spamalytics}
\bibfield{author}{\bibinfo{person}{Chris Kanich}, \bibinfo{person}{Christian
  Kreibich}, \bibinfo{person}{Kirill Levchenko}, \bibinfo{person}{Brandon
  Enright}, \bibinfo{person}{Geoffrey~M Voelker}, \bibinfo{person}{Vern
  Paxson}, {and} \bibinfo{person}{Stefan Savage}.}
  \bibinfo{year}{2008}\natexlab{}.
\newblock \showarticletitle{Spamalytics: An empirical analysis of spam
  marketing conversion}. In \bibinfo{booktitle}{\emph{15th ACM conference on
  Computer and communications security}}.
\newblock


\bibitem[\protect\citeauthoryear{Kiene, Monroy-Hern\'{a}ndez, and Hill}{Kiene
  et~al\mbox{.}}{2016}]%
        {kiene2016surviving}
\bibfield{author}{\bibinfo{person}{Charles Kiene}, \bibinfo{person}{Andr{\'e}s
  Monroy-Hern\'{a}ndez}, {and} \bibinfo{person}{Benjamin~Mako Hill}.}
  \bibinfo{year}{2016}\natexlab{}.
\newblock \showarticletitle{Surviving an ``Eternal September'': How an Online
  Community Managed a Surge of Newcomers}. In
  \bibinfo{booktitle}{\emph{Proceedings of the 2016 CHI Conference on Human
  Factors in Computing Systems}} (San Jose, California, USA)
  \emph{(\bibinfo{series}{CHI '16})}. \bibinfo{publisher}{ACM},
  \bibinfo{address}{New York, NY, USA}, \bibinfo{pages}{1152--1156}.
\newblock
\showISBNx{978-1-4503-3362-7}
\urldef\tempurl%
\url{https://doi.org/10.1145/2858036.2858356}
\showDOI{\tempurl}


\bibitem[\protect\citeauthoryear{Kumar, Cheng, Leskovec, and
  Subrahmanian}{Kumar et~al\mbox{.}}{2017}]%
        {kumar2017sockpuppet}
\bibfield{author}{\bibinfo{person}{Srijan Kumar}, \bibinfo{person}{Justin
  Cheng}, \bibinfo{person}{Jure Leskovec}, {and} \bibinfo{person}{V.S.
  Subrahmanian}.} \bibinfo{year}{2017}\natexlab{}.
\newblock \showarticletitle{An Army of Me: Sockpuppets in Online Discussion
  Communities}. In \bibinfo{booktitle}{\emph{WWW '17}}.
\newblock


\bibitem[\protect\citeauthoryear{Ling, Balci, Blackburn, and Stringhini}{Ling
  et~al\mbox{.}}{2021}]%
        {ling2021zoom}
\bibfield{author}{\bibinfo{person}{Chen Ling}, \bibinfo{person}{Utkucan Balci},
  \bibinfo{person}{Jeremy Blackburn}, {and} \bibinfo{person}{Gianluca
  Stringhini}.} \bibinfo{year}{2021}\natexlab{}.
\newblock \showarticletitle{A First Look at Zoombombing}. In
  \bibinfo{booktitle}{\emph{IEEESP}}.
\newblock


\bibitem[\protect\citeauthoryear{MacKinnon}{MacKinnon}{1997}]%
        {mackinnon1997virtual}
\bibfield{author}{\bibinfo{person}{Richard MacKinnon}.}
  \bibinfo{year}{1997}\natexlab{}.
\newblock \showarticletitle{Virtual Rape}.
\newblock \bibinfo{journal}{\emph{Journal of Computer-Mediated Communication}}
  \bibinfo{volume}{2}, \bibinfo{number}{4} (\bibinfo{year}{1997}),
  \bibinfo{pages}{1--2}.
\newblock
\urldef\tempurl%
\url{https://doi.org/10.1111/j.1083-6101.1997.tb00200.x}
\showDOI{\tempurl}


\bibitem[\protect\citeauthoryear{Mahar, Zhang, and Karger}{Mahar
  et~al\mbox{.}}{2018}]%
        {mahar2018squadbox}
\bibfield{author}{\bibinfo{person}{Kaitlin Mahar}, \bibinfo{person}{Amy~X.
  Zhang}, {and} \bibinfo{person}{David Karger}.}
  \bibinfo{year}{2018}\natexlab{}.
\newblock \showarticletitle{Squadbox: A Tool to Combat Email Harassment Using
  Friendsourced Moderation}. In \bibinfo{booktitle}{\emph{Proceedings of the
  2018 CHI Conference on Human Factors in Computing Systems}} (Montreal QC,
  Canada) \emph{(\bibinfo{series}{CHI '18})}. \bibinfo{publisher}{ACM},
  \bibinfo{address}{New York, NY, USA}, Article \bibinfo{articleno}{586},
  \bibinfo{numpages}{13}~pages.
\newblock
\showISBNx{978-1-4503-5620-6}
\urldef\tempurl%
\url{https://doi.org/10.1145/3173574.3174160}
\showDOI{\tempurl}


\bibitem[\protect\citeauthoryear{Mariconti, Suarez-Tangil, Blackburn,
  De~Cristofaro, Kourtellis, Leontiadis, Serrano, and Stringhini}{Mariconti
  et~al\mbox{.}}{[n.d.]}]%
        {mariconti2019you}
\bibfield{author}{\bibinfo{person}{Enrico Mariconti},
  \bibinfo{person}{Guillermo Suarez-Tangil}, \bibinfo{person}{Jeremy
  Blackburn}, \bibinfo{person}{Emiliano De~Cristofaro},
  \bibinfo{person}{Nicolas Kourtellis}, \bibinfo{person}{Ilias Leontiadis},
  \bibinfo{person}{Jordi~Luque Serrano}, {and} \bibinfo{person}{Gianluca
  Stringhini}.} \bibinfo{year}{[n.d.]}\natexlab{}.
\newblock \showarticletitle{"You Know What to Do" Proactive Detection of
  YouTube Videos Targeted by Coordinated Hate Attacks}.
\newblock \bibinfo{journal}{\emph{Proceedings of the ACM on Human-Computer
  Interaction}} \bibinfo{number}{CSCW} (\bibinfo{year}{[n.\,d.]}).
\newblock


\bibitem[\protect\citeauthoryear{Marwick}{Marwick}{2021}]%
        {marwick2021networked}
\bibfield{author}{\bibinfo{person}{Alice~E Marwick}.}
  \bibinfo{year}{2021}\natexlab{}.
\newblock \showarticletitle{{Morally Motivated Networked Harassment as
  Normative Reinforcement}}.
\newblock \bibinfo{journal}{\emph{Social Media + Society}} \bibinfo{volume}{7},
  \bibinfo{number}{2} (\bibinfo{year}{2021}),
  \bibinfo{pages}{20563051211021378}.
\newblock
\urldef\tempurl%
\url{https://doi.org/10.1177/20563051211021378}
\showDOI{\tempurl}


\bibitem[\protect\citeauthoryear{Massanari}{Massanari}{2017}]%
        {massanari2017gamergate}
\bibfield{author}{\bibinfo{person}{Adrienne Massanari}.}
  \bibinfo{year}{2017}\natexlab{}.
\newblock \showarticletitle{{{\#}Gamergate and The Fappening: How Reddit's
  algorithm, governance, and culture support toxic technocultures}}.
\newblock \bibinfo{journal}{\emph{New Media {\&} Society}}
  \bibinfo{volume}{19}, \bibinfo{number}{3} (\bibinfo{year}{2017}),
  \bibinfo{pages}{329--346}.
\newblock


\bibitem[\protect\citeauthoryear{Metaxa-Kakavouli, Maas, and
  Aldrich}{Metaxa-Kakavouli et~al\mbox{.}}{2018}]%
        {metaxa2018hurricane}
\bibfield{author}{\bibinfo{person}{Dana\"{e} Metaxa-Kakavouli},
  \bibinfo{person}{Paige Maas}, {and} \bibinfo{person}{Daniel~P. Aldrich}.}
  \bibinfo{year}{2018}\natexlab{}.
\newblock \showarticletitle{How Social Ties Influence Hurricane Evacuation
  Behavior}.
\newblock \bibinfo{journal}{\emph{Proc. ACM Hum.-Comput. Interact.}}
  \bibinfo{volume}{2}, \bibinfo{number}{CSCW}, Article \bibinfo{articleno}{122}
  (\bibinfo{date}{nov} \bibinfo{year}{2018}), \bibinfo{numpages}{16}~pages.
\newblock
\urldef\tempurl%
\url{https://doi.org/10.1145/3274391}
\showDOI{\tempurl}


\bibitem[\protect\citeauthoryear{Monroy-Hern\'{a}ndez, boyd, Kiciman,
  De~Choudhury, and Counts}{Monroy-Hern\'{a}ndez et~al\mbox{.}}{2013}]%
        {monroyhernandez2013correspondents}
\bibfield{author}{\bibinfo{person}{Andr\'{e}s Monroy-Hern\'{a}ndez},
  \bibinfo{person}{danah boyd}, \bibinfo{person}{Emre Kiciman},
  \bibinfo{person}{Munmun De~Choudhury}, {and} \bibinfo{person}{Scott Counts}.}
  \bibinfo{year}{2013}\natexlab{}.
\newblock \showarticletitle{The New War Correspondents: The Rise of Civic Media
  Curation in Urban Warfare}. In \bibinfo{booktitle}{\emph{Proceedings of the
  2013 Conference on Computer Supported Cooperative Work}} (San Antonio, Texas,
  USA) \emph{(\bibinfo{series}{CSCW '13})}. \bibinfo{publisher}{Association for
  Computing Machinery}, \bibinfo{address}{New York, NY, USA},
  \bibinfo{pages}{1443–1452}.
\newblock
\showISBNx{9781450313315}
\urldef\tempurl%
\url{https://doi.org/10.1145/2441776.2441938}
\showDOI{\tempurl}


\bibitem[\protect\citeauthoryear{Moore, Shannon, Brown, Voelker, and
  Savage}{Moore et~al\mbox{.}}{2006}]%
        {moore2006inferring}
\bibfield{author}{\bibinfo{person}{David Moore}, \bibinfo{person}{Colleen
  Shannon}, \bibinfo{person}{Douglas~J Brown}, \bibinfo{person}{Geoffrey~M
  Voelker}, {and} \bibinfo{person}{Stefan Savage}.}
  \bibinfo{year}{2006}\natexlab{}.
\newblock \showarticletitle{Inferring internet denial-of-service activity}.
\newblock \bibinfo{journal}{\emph{ACM Transactions on Computer Systems (TOCS)}}
  (\bibinfo{year}{2006}).
\newblock


\bibitem[\protect\citeauthoryear{Noble}{Noble}{2018}]%
        {Noble2018algorithms}
\bibfield{author}{\bibinfo{person}{Safiya~Umoja Noble}.}
  \bibinfo{year}{2018}\natexlab{}.
\newblock \bibinfo{booktitle}{\emph{{Algorithms of Oppression}}}.
\newblock \bibinfo{publisher}{NYU Press}, \bibinfo{address}{New York, NY, USA}.
\newblock
\urldef\tempurl%
\url{https://doi.org/10.2307/j.ctt1pwt9w5}
\showDOI{\tempurl}


\bibitem[\protect\citeauthoryear{Palen and Anderson}{Palen and
  Anderson}{2016}]%
        {palen2016newdata}
\bibfield{author}{\bibinfo{person}{Leysia Palen} {and}
  \bibinfo{person}{Kenneth~M Anderson}.} \bibinfo{year}{2016}\natexlab{}.
\newblock \showarticletitle{{Crisis informatics--New data for extraordinary
  times}}.
\newblock \bibinfo{journal}{\emph{Science}} \bibinfo{volume}{353},
  \bibinfo{number}{6296} (\bibinfo{year}{2016}), \bibinfo{pages}{224--225}.
\newblock
\urldef\tempurl%
\url{https://doi.org/10.1126/science.aag2579}
\showDOI{\tempurl}


\bibitem[\protect\citeauthoryear{Palen, Vieweg, Sutton, Liu, and Hughes}{Palen
  et~al\mbox{.}}{2007}]%
        {palen2007crisis}
\bibfield{author}{\bibinfo{person}{Leysia Palen}, \bibinfo{person}{Sarah
  Vieweg}, \bibinfo{person}{Jeannette Sutton}, \bibinfo{person}{Sophia~B Liu},
  {and} \bibinfo{person}{Amanda Hughes}.} \bibinfo{year}{2007}\natexlab{}.
\newblock \showarticletitle{Crisis informatics: Studying crisis in a networked
  world}. In \bibinfo{booktitle}{\emph{Proceedings of the Third International
  Conference on E-Social Science}}. \bibinfo{pages}{7--9}.
\newblock


\bibitem[\protect\citeauthoryear{Pandey}{Pandey}{[n.d.]}]%
        {pandey2021NEWShateraids}
\bibfield{author}{\bibinfo{person}{Manish Pandey}.}
  \bibinfo{year}{[n.d.]}\natexlab{}.
\newblock \showarticletitle{Twitch announces new tools to fight hate raids}.
\newblock \bibinfo{journal}{\emph{BBC}} (\bibinfo{year}{[n.\,d.]}).
\newblock
\urldef\tempurl%
\url{https://www.bbc.com/news/newsbeat-58594732}
\showURL{%
\tempurl}


\bibitem[\protect\citeauthoryear{Parrish}{Parrish}{[n.d.]}]%
        {parrish2021NEWSprotests}
\bibfield{author}{\bibinfo{person}{Ash Parrish}.}
  \bibinfo{year}{[n.d.]}\natexlab{}.
\newblock \showarticletitle{Twitch viewership noticeably dropped when streamers
  took a day off in protest}.
\newblock \bibinfo{journal}{\emph{The Verge}} (\bibinfo{year}{[n.\,d.]}).
\newblock
\urldef\tempurl%
\url{https://www.theverge.com/2021/9/2/22654534/streamers-twitch-walkout-viewership-drop}
\showURL{%
\tempurl}


\bibitem[\protect\citeauthoryear{Phillips}{Phillips}{2011}]%
        {Phillips2011memorial}
\bibfield{author}{\bibinfo{person}{Whitney Phillips}.}
  \bibinfo{year}{2011}\natexlab{}.
\newblock \showarticletitle{{LOLing at tragedy: Facebook trolls, memorial pages
  and resistance to grief online}}.
\newblock \bibinfo{journal}{\emph{First Monday}} \bibinfo{volume}{16},
  \bibinfo{number}{12} (\bibinfo{year}{2011}), \bibinfo{numpages}{12}~pages.
\newblock
\showISBNx{1396-0466, 1396-0466}
\showISSN{13960466}
\urldef\tempurl%
\url{https://doi.org/10.5210/fm.v16i12.3168}
\showDOI{\tempurl}


\bibitem[\protect\citeauthoryear{Phillips}{Phillips}{2015}]%
        {phillips2015we}
\bibfield{author}{\bibinfo{person}{Whitney Phillips}.}
  \bibinfo{year}{2015}\natexlab{}.
\newblock \bibinfo{booktitle}{\emph{This is why we can't have nice things:
  Mapping the relationship between online trolling and mainstream culture}}.
\newblock \bibinfo{publisher}{MIT Press}, \bibinfo{address}{Cambridge, MA,
  USA}.
\newblock


\bibitem[\protect\citeauthoryear{Phillips}{Phillips}{2019}]%
        {phillips2019exclusionary}
\bibfield{author}{\bibinfo{person}{Whitney Phillips}.}
  \bibinfo{year}{2019}\natexlab{}.
\newblock \showarticletitle{{It Wasn't Just the Trolls: Early Internet Culture,
  “Fun,” and the Fires of Exclusionary Laughter}}.
\newblock \bibinfo{journal}{\emph{Social Media + Society}} \bibinfo{volume}{5},
  \bibinfo{number}{3} (\bibinfo{year}{2019}),
  \bibinfo{pages}{2056305119849493}.
\newblock
\urldef\tempurl%
\url{https://doi.org/10.1177/2056305119849493}
\showDOI{\tempurl}


\bibitem[\protect\citeauthoryear{Polhamus}{Polhamus}{[n.d.]}]%
        {Polhamus2022NEWShateraids}
\bibfield{author}{\bibinfo{person}{Blaine Polhamus}.}
  \bibinfo{year}{[n.d.]}\natexlab{}.
\newblock \showarticletitle{Hate raids return to Twitch, another wave of
  attacks target LGBTQIA+ streamers}.
\newblock \bibinfo{journal}{\emph{DOT ESPORTS}} (\bibinfo{year}{[n.\,d.]}).
\newblock
\urldef\tempurl%
\url{https://dotesports.com/streaming/news/hate-raids-return-to-twitch-another-wave-of-attacks-target-lgbtqia-streamers}
\showURL{%
\tempurl}


\bibitem[\protect\citeauthoryear{Roberts}{Roberts}{2016}]%
        {roberts2016commercial}
\bibfield{author}{\bibinfo{person}{Sarah~T. Roberts}.}
  \bibinfo{year}{2016}\natexlab{}.
\newblock \showarticletitle{{Commercial Content Moderation: Digital Laborers'
  Dirty Work}}.
\newblock In \bibinfo{booktitle}{\emph{The Intersectional Internet: Race, Sex,
  Class and Culture Online}}, \bibfield{editor}{\bibinfo{person}{Safiya~Umoja
  Noble} {and} \bibinfo{person}{Brendesha~M. Tynes}} (Eds.).
  \bibinfo{publisher}{Peter Lang Digital Formations series},
  \bibinfo{address}{New York, NY, USA}, \bibinfo{pages}{147--160}.
\newblock


\bibitem[\protect\citeauthoryear{Scheuerman, Branham, and Hamidi}{Scheuerman
  et~al\mbox{.}}{2018}]%
        {scheuerman2018safespaces}
\bibfield{author}{\bibinfo{person}{Morgan~Klaus Scheuerman},
  \bibinfo{person}{Stacy~M. Branham}, {and} \bibinfo{person}{Foad Hamidi}.}
  \bibinfo{year}{2018}\natexlab{}.
\newblock \showarticletitle{Safe Spaces and Safe Places: Unpacking
  Technology-Mediated Experiences of Safety and Harm with Transgender People}.
\newblock \bibinfo{journal}{\emph{Proc. ACM Hum.-Comput. Interact.}}
  \bibinfo{volume}{2}, \bibinfo{number}{CSCW}, Article \bibinfo{articleno}{155}
  (\bibinfo{date}{Nov.} \bibinfo{year}{2018}), \bibinfo{numpages}{27}~pages.
\newblock
\urldef\tempurl%
\url{https://doi.org/10.1145/3274424}
\showDOI{\tempurl}


\bibitem[\protect\citeauthoryear{Seering, Dym, Kaufman, and Bernstein}{Seering
  et~al\mbox{.}}{2022}]%
        {seering2022pride}
\bibfield{author}{\bibinfo{person}{Joseph Seering}, \bibinfo{person}{Brianna
  Dym}, \bibinfo{person}{Geoff Kaufman}, {and} \bibinfo{person}{Michael
  Bernstein}.} \bibinfo{year}{2022}\natexlab{}.
\newblock \showarticletitle{Pride and Professionalization in Volunteer
  Moderation: Lessons for Effective in Platform-User Collaboration}.
\newblock \bibinfo{journal}{\emph{Journal of Online Trust and Safety}}
  (\bibinfo{year}{2022}).
\newblock


\bibitem[\protect\citeauthoryear{Seering and Kairam}{Seering and
  Kairam}{2022}]%
        {seering2022whomoderates}
\bibfield{author}{\bibinfo{person}{Joseph Seering} {and}
  \bibinfo{person}{Sanjay~R. Kairam}.} \bibinfo{year}{2022}\natexlab{}.
\newblock \showarticletitle{Who Moderates on Twitch and What Do They Do?
  Quantifying Practices in Community Moderation on Twitch}.
\newblock \bibinfo{journal}{\emph{Proc. ACM Hum.-Comput. Interact.}}
  \bibinfo{volume}{7}, \bibinfo{number}{GROUP}, Article \bibinfo{articleno}{18}
  (\bibinfo{date}{dec} \bibinfo{year}{2022}), \bibinfo{numpages}{18}~pages.
\newblock
\urldef\tempurl%
\url{https://doi.org/10.1145/3567568}
\showDOI{\tempurl}


\bibitem[\protect\citeauthoryear{Seering, Kraut, and Dabbish}{Seering
  et~al\mbox{.}}{2017}]%
        {Seering2017shaping}
\bibfield{author}{\bibinfo{person}{Joseph Seering}, \bibinfo{person}{Robert
  Kraut}, {and} \bibinfo{person}{Laura Dabbish}.}
  \bibinfo{year}{2017}\natexlab{}.
\newblock \showarticletitle{Shaping Pro and Anti-Social Behavior on Twitch
  Through Moderation and Example-Setting}. In
  \bibinfo{booktitle}{\emph{Proceedings of the 2017 ACM Conference on Computer
  Supported Cooperative Work and Social Computing}} (Portland, Oregon, USA)
  \emph{(\bibinfo{series}{CSCW '17})}. \bibinfo{publisher}{ACM},
  \bibinfo{address}{New York, NY, USA}, \bibinfo{pages}{111--125}.
\newblock
\showISBNx{978-1-4503-4335-0}
\urldef\tempurl%
\url{https://doi.org/10.1145/2998181.2998277}
\showDOI{\tempurl}


\bibitem[\protect\citeauthoryear{Seering, Wang, Yoon, and Kaufman}{Seering
  et~al\mbox{.}}{2019}]%
        {seering2019moderator}
\bibfield{author}{\bibinfo{person}{Joseph Seering}, \bibinfo{person}{Tony
  Wang}, \bibinfo{person}{Jina Yoon}, {and} \bibinfo{person}{Geoff Kaufman}.}
  \bibinfo{year}{2019}\natexlab{}.
\newblock \showarticletitle{Moderator engagement and community development in
  the age of algorithms}.
\newblock \bibinfo{journal}{\emph{New Media \& Society}} \bibinfo{volume}{21},
  \bibinfo{number}{7} (\bibinfo{year}{2019}), \bibinfo{pages}{1417--1443}.
\newblock
\urldef\tempurl%
\url{https://doi.org/10.1177/1461444818821316}
\showDOI{\tempurl}


\bibitem[\protect\citeauthoryear{Semaan and Mark}{Semaan and Mark}{2011}]%
        {semaan2011breakdowns}
\bibfield{author}{\bibinfo{person}{Bryan Semaan} {and} \bibinfo{person}{Gloria
  Mark}.} \bibinfo{year}{2011}\natexlab{}.
\newblock \showarticletitle{Technology-Mediated Social Arrangements to Resolve
  Breakdowns in Infrastructure during Ongoing Disruption}.
\newblock \bibinfo{journal}{\emph{ACM Trans. Comput.-Hum. Interact.}}
  \bibinfo{volume}{18}, \bibinfo{number}{4}, Article \bibinfo{articleno}{21}
  (\bibinfo{date}{12} \bibinfo{year}{2011}), \bibinfo{numpages}{21}~pages.
\newblock
\showISSN{1073-0516}
\urldef\tempurl%
\url{https://doi.org/10.1145/2063231.2063235}
\showDOI{\tempurl}


\bibitem[\protect\citeauthoryear{Smith}{Smith}{1999}]%
        {Smith1999communities}
\bibfield{author}{\bibinfo{person}{Anna~DuVal Smith}.}
  \bibinfo{year}{1999}\natexlab{}.
\newblock \showarticletitle{{Problems of Conflict Management in Virtual
  Communities}}.
\newblock In \bibinfo{booktitle}{\emph{Communities in Cyberspace}
  (\bibinfo{edition}{1st} ed.)}, \bibfield{editor}{\bibinfo{person}{Marc~A
  Smith} {and} \bibinfo{person}{P~Kollock}} (Eds.).
  \bibinfo{publisher}{Routledge}, \bibinfo{address}{New York, NY, USA},
  \bibinfo{pages}{135--166}.
\newblock


\bibitem[\protect\citeauthoryear{Soden and Palen}{Soden and Palen}{2014}]%
        {soden2014crowdsourced}
\bibfield{author}{\bibinfo{person}{Robert Soden} {and} \bibinfo{person}{Leysia
  Palen}.} \bibinfo{year}{2014}\natexlab{}.
\newblock \showarticletitle{{From Crowdsourced Mapping to Community Mapping:
  The Post-earthquake Work of OpenStreetMap Haiti}}. In
  \bibinfo{booktitle}{\emph{COOP 2014 - Proceedings of the 11th International
  Conference on the Design of Cooperative Systems, 27-30 May 2014, Nice
  (France)}}, \bibfield{editor}{\bibinfo{person}{Chiara Rossitto},
  \bibinfo{person}{Luigina Ciolfi}, \bibinfo{person}{David Martin}, {and}
  \bibinfo{person}{Bernard Conein}} (Eds.). \bibinfo{publisher}{Springer
  International Publishing}, \bibinfo{address}{Cham},
  \bibinfo{pages}{311--326}.
\newblock
\showISBNx{978-3-319-06498-7}


\bibitem[\protect\citeauthoryear{Soden and Palen}{Soden and Palen}{2016}]%
        {soden2016infrastructure}
\bibfield{author}{\bibinfo{person}{Robert Soden} {and} \bibinfo{person}{Leysia
  Palen}.} \bibinfo{year}{2016}\natexlab{}.
\newblock \showarticletitle{Infrastructure in the Wild: What Mapping in
  Post-Earthquake Nepal Reveals about Infrastructural Emergence}. In
  \bibinfo{booktitle}{\emph{Proceedings of the 2016 CHI Conference on Human
  Factors in Computing Systems}} (San Jose, California, USA)
  \emph{(\bibinfo{series}{CHI '16})}. \bibinfo{publisher}{Association for
  Computing Machinery}, \bibinfo{address}{New York, NY, USA},
  \bibinfo{pages}{2796–2807}.
\newblock
\showISBNx{9781450333627}
\urldef\tempurl%
\url{https://doi.org/10.1145/2858036.2858545}
\showDOI{\tempurl}


\bibitem[\protect\citeauthoryear{Stephen}{Stephen}{[n.d.]}]%
        {stephen202NEWSlivestreaming}
\bibfield{author}{\bibinfo{person}{Bijan Stephen}.}
  \bibinfo{year}{[n.d.]}\natexlab{}.
\newblock \showarticletitle{The lockdown live-streaming numbers are out, and
  they’re huge}.
\newblock \bibinfo{journal}{\emph{The Verge}} (\bibinfo{year}{[n.\,d.]}).
\newblock
\urldef\tempurl%
\url{https://www.theverge.com/2020/5/13/21257227/coronavirus-streamelements-arsenalgg-twitch-youtube-livestream-numbers}
\showURL{%
\tempurl}


\bibitem[\protect\citeauthoryear{Thach, Mayworm, Delmonaco, and Haimson}{Thach
  et~al\mbox{.}}{2022}]%
        {thach2022invisible}
\bibfield{author}{\bibinfo{person}{Hibby Thach}, \bibinfo{person}{Samuel
  Mayworm}, \bibinfo{person}{Daniel Delmonaco}, {and} \bibinfo{person}{Oliver
  Haimson}.} \bibinfo{year}{2022}\natexlab{}.
\newblock \showarticletitle{(In)visible moderation: A digital ethnography of
  marginalized users and content moderation on Twitch and Reddit}.
\newblock \bibinfo{journal}{\emph{New Media {\&} Society}}
  (\bibinfo{year}{2022}).
\newblock


\bibitem[\protect\citeauthoryear{Thomas, Akhawe, Bailey, Boneh, Consolvo, Dell,
  Durumeric, Kelley, Kumar, McCoy, Meiklejohn, Ristenpart, and
  Stringhini}{Thomas et~al\mbox{.}}{2021}]%
        {thomas2021hatesok}
\bibfield{author}{\bibinfo{person}{Kurt Thomas}, \bibinfo{person}{Devdatta
  Akhawe}, \bibinfo{person}{Michael Bailey}, \bibinfo{person}{Dan Boneh},
  \bibinfo{person}{Sunny Consolvo}, \bibinfo{person}{Nicola Dell},
  \bibinfo{person}{Zakir Durumeric}, \bibinfo{person}{Patrick~Gage Kelley},
  \bibinfo{person}{Deepak Kumar}, \bibinfo{person}{Damon McCoy},
  \bibinfo{person}{Sarah Meiklejohn}, \bibinfo{person}{Thomas Ristenpart},
  {and} \bibinfo{person}{Gianluca Stringhini}.}
  \bibinfo{year}{2021}\natexlab{}.
\newblock \showarticletitle{SoK: Hate, Harassment, and the Changing Landscape
  of Online Abuse}. In \bibinfo{booktitle}{\emph{IEEESP}}.
\newblock


\bibitem[\protect\citeauthoryear{Thomas, Grier, Song, and Paxson}{Thomas
  et~al\mbox{.}}{2011}]%
        {thomas2011suspended}
\bibfield{author}{\bibinfo{person}{Kurt Thomas}, \bibinfo{person}{Chris Grier},
  \bibinfo{person}{Dawn Song}, {and} \bibinfo{person}{Vern Paxson}.}
  \bibinfo{year}{2011}\natexlab{}.
\newblock \showarticletitle{Suspended accounts in retrospect: an analysis of
  twitter spam}. In \bibinfo{booktitle}{\emph{ACM SIGCOMM conference on
  Internet measurement conference}}.
\newblock


\bibitem[\protect\citeauthoryear{Thomas, Iatskiv, Bursztein, Pietraszek, Grier,
  and McCoy}{Thomas et~al\mbox{.}}{2014}]%
        {thomas2014dialing}
\bibfield{author}{\bibinfo{person}{Kurt Thomas}, \bibinfo{person}{Dmytro
  Iatskiv}, \bibinfo{person}{Elie Bursztein}, \bibinfo{person}{Tadek
  Pietraszek}, \bibinfo{person}{Chris Grier}, {and} \bibinfo{person}{Damon
  McCoy}.} \bibinfo{year}{2014}\natexlab{}.
\newblock \showarticletitle{Dialing back abuse on phone verified accounts}. In
  \bibinfo{booktitle}{\emph{ACM SIGSAC Conference on Computer and
  Communications Security}}.
\newblock


\bibitem[\protect\citeauthoryear{Twitch~Interactive}{Twitch~Interactive}{[n.d.]}]%
        {twitch2021NEWStags}
\bibfield{author}{\bibinfo{person}{Inc. Twitch~Interactive}.}
  \bibinfo{year}{[n.d.]}\natexlab{}.
\newblock \showarticletitle{Celebrate Yourself and Your Community with 350+ New
  Tags}.
\newblock \bibinfo{journal}{\emph{Twitch Blog}} (\bibinfo{year}{[n.\,d.]}).
\newblock
\urldef\tempurl%
\url{https://blog.twitch.tv/en/2021/05/26/celebrate-yourself-and-your-community-with-350-new-tags/}
\showURL{%
\tempurl}


\bibitem[\protect\citeauthoryear{Vitak, Chadha, Steiner, and Ashktorab}{Vitak
  et~al\mbox{.}}{2017}]%
        {vitak2017identifying}
\bibfield{author}{\bibinfo{person}{Jessica Vitak}, \bibinfo{person}{Kalyani
  Chadha}, \bibinfo{person}{Linda Steiner}, {and} \bibinfo{person}{Zahra
  Ashktorab}.} \bibinfo{year}{2017}\natexlab{}.
\newblock \showarticletitle{Identifying Women's Experiences With and Strategies
  for Mitigating Negative Effects of Online Harassment}. In
  \bibinfo{booktitle}{\emph{Proceedings of the 2017 ACM Conference on Computer
  Supported Cooperative Work and Social Computing}} (Portland, Oregon, USA)
  \emph{(\bibinfo{series}{CSCW '17})}. \bibinfo{publisher}{ACM},
  \bibinfo{address}{New York, NY, USA}, \bibinfo{pages}{1231--1245}.
\newblock
\showISBNx{978-1-4503-4335-0}
\urldef\tempurl%
\url{https://doi.org/10.1145/2998181.2998337}
\showDOI{\tempurl}


\bibitem[\protect\citeauthoryear{Yosilewitz}{Yosilewitz}{[n.d.]}]%
        {yosilewitz2021NEWSstats}
\bibfield{author}{\bibinfo{person}{Adam Yosilewitz}.}
  \bibinfo{year}{[n.d.]}\natexlab{}.
\newblock \showarticletitle{StreamElements Analysis on Twitch Bullying}.
\newblock \bibinfo{journal}{\emph{StreamElements}} (\bibinfo{year}{[n.\,d.]}).
\newblock
\urldef\tempurl%
\url{https://blog.streamelements.com/streamelements-analysis-on-twitch-bullying-c3f2b2240318}
\showURL{%
\tempurl}


\bibitem[\protect\citeauthoryear{Zade, Shah, Rangarajan, Kshirsagar, Imran, and
  Starbird}{Zade et~al\mbox{.}}{2018}]%
        {zade2018crisisresponse}
\bibfield{author}{\bibinfo{person}{Himanshu Zade}, \bibinfo{person}{Kushal
  Shah}, \bibinfo{person}{Vaibhavi Rangarajan}, \bibinfo{person}{Priyanka
  Kshirsagar}, \bibinfo{person}{Muhammad Imran}, {and} \bibinfo{person}{Kate
  Starbird}.} \bibinfo{year}{2018}\natexlab{}.
\newblock \showarticletitle{From Situational Awareness to Actionability:
  Towards Improving the Utility of Social Media Data for Crisis Response}.
\newblock \bibinfo{journal}{\emph{Proc. ACM Hum.-Comput. Interact.}}
  \bibinfo{volume}{2}, \bibinfo{number}{CSCW}, Article \bibinfo{articleno}{195}
  (\bibinfo{date}{nov} \bibinfo{year}{2018}), \bibinfo{numpages}{18}~pages.
\newblock
\urldef\tempurl%
\url{https://doi.org/10.1145/3274464}
\showDOI{\tempurl}


\end{thebibliography}
\received{July 2022}
\received[revised]{October 2022}
\received[accepted]{January 2023}
\appendix
\section{Primary Interview Questions}
\label{appendix:interview_questions}
\begin{enumerate}
    \item Over the course of the past few months, have you been impacted either directly or indirectly by hate raids?
    \begin{enumerate}
        \item If so, how?
        \item (If they were hate raided or observed hate raids) Can you describe what the hate raid(s) were like?
    \end{enumerate}
    
    \item What did you do to protect yourself from hate raids if anything?
    \begin{enumerate}
        \item (If not mentioned) Did you add any new moderation tools?
        \item (If not mentioned) Did you add new moderators or request additional moderator support?
    \end{enumerate}
    
    \item How effective were each of these strategies for protecting yourself?
    
    \item How did you learn about new ways to protect yourself?
        \begin{enumerate}
            \item (If not mentioned) How did you learn about how to use new tools?
            \item (If not mentioned) How did you learn about any new strategies for protecting your personal information?
            \item (If not mentioned) Were you involved in any spaces were these topics were discussed?
        \end{enumerate}
        
    \item Were you involved in any forms of collective action like \#TwitchDoBetter or \#ADayOffTwitch?

    \item How do you feel about Twitch’s response to the Hate Raids?
    \begin{enumerate}
        \item How do you feel about the lawsuit that Twitch has announced against the perpetrators?
        \item Do you think that the new moderation features Twitch released have helped?
        \item How do you feel about the way Twitch communicated about the Hate Raids in August and September?
    \end{enumerate}
    
    \item What do you think Twitch could do better in the future in handling cases like this one?
\end{enumerate}

\section{Interview Codebooks}
\label{interview_codebooks}

\begin{table}[h]
\small
\setlength{\tabcolsep}{1.5pt}
\centering
\begin{tabularx}{.95\columnwidth}{XX}
\toprule
Streamer Category Label & Description\\
\midrule
Effectiveness of Twitch's responses &  Chunks about specific aspects of Twitch’s responses, including communication, lawsuit, tools they added, etc. \\
Instrumental community support      &  Chunks about community resource sharing and instrumental/informational support \\
Social community support            &  Chunks about social/interpersonal support they received \\
Community organization              &  Chunks about collective action or group-organized things, e.g., \#TwitchDoBetter \\
Degree of raid targeting            &  Degree of attack personalization (how targeted?) \\
Frequency of hate raids experienced &  Frequency (how often?) \\
Raid vectors                        &  Different attack vectors (how many different ways) \\
Raid responses (short-term)         &  Things the streamer did in-the-moment or during the weeks while the hate raids were going on to protect themselves/others \\
Raid impact (long-term)             &  Longer-term impact on streamers’ careers, health, well-being \\
\bottomrule
\end{tabularx}

\begin{tabularx}{.95\columnwidth}{XX}
\toprule
Bot Developer Category Label & Description\\
\midrule
Community need                   &  Chunks about how a need for specific third-party resources for the community revealed, if streamers ask specifically for features, developers’ own observations/pro-social motivations, and gaps/shortcomings in Twitch-provided tools \\
Developer dependence             &  Chunks about the degree to which streamers (large vs. small might have different experiences) depend on third-party bot developers to better protect themselves from hate raids, how many channels (how
was the adoption), how effective (numbers of raids/bots/messages intercepted or moderated) \\
Hate raid arms race              &  Chunks about the kind of arms race or ``cat and mouse game'' bot developers experienced while rolling out features to combat hate raids \\
Effectiveness of Twitch tools    &  Chunks about bot developers’ perspectives on the efficacy of Twitch’s technical tools before/after the hate raids \\
Twitch development obstacles     &  Chunks about Twitch obstacles/hurdles that made effective bot development difficult \\
Twitch communication             &  Chunks about Twitch communications with the community (and how it fueled their dissatisfaction with the platform) \\
Developer coordination           &  Chunks about how the community of developers organized \\
\bottomrule
\end{tabularx}
\end{table}
\end{document}